# Resonantly Enhanced Multiphonon Scattering and Local Orbital-Phonon Coupling in Bulk and Thin Flakes of 2D Single Crystals of Antiferromagnetic $(Ni_xFe_{1-x})_2P_2S_6$

Nasaru Khan[1*], Miłosz Rybak[2], Yuliia Shemerliuk[3], Sebastian Selter[3], Bernd Büchner[3,4], Saicharan Aswartham[3,5], Krzysztof Wohlfeld[6] and Pradeep Kumar[1,†]

[1] *School of Physical Sciences, Indian Institute of Technology Mandi, Mandi-175005, India*
[2]*Department of Semiconductor Materials Engineering, Faculty of Fundamental Problems of Technology, Wrocław University of Science and Technology, Wybrzeże Wyspiańskiego 27, 50-370 Wrocław, Poland*
[3]*Leibniz-Institute for Solid-state and Materials Research, IFW-Dresden, 01069 Dresden, Germany*
[4]*Institute of Solid State and Materials Physics and Würzburg-Dresden Cluster of Excellence ct.qmat, Technische Universität Dresden, 01062 Dresden, Germany*
[5]*International Research Centre MagTop, Institute of Physics, Polish Academy of Sciences, al. Lotników 32/46, 02-668 Warsaw, Poland*
[6]*Institute of Theoretical Physics, Faculty of Physics, University of Warsaw, Pasteura St. 5, 02-093 Warsaw, Poland*

## Abstract

Orbital degrees of freedom play a pivotal role in shaping the physical properties of two-dimensional (2D) van der Waals magnetic systems, strongly influencing electron-phonon coupling and intermediate-state dynamics. In this study, we present a comprehensive temperature and thickness-dependent Raman investigation of the single crystals of 2D van der Waals antiferromagnetic series $(Ni_xFe_{1-x})_2P_2S_6$. Utilizing Raman spectroscopy, we explore the interplay between local orbital excitations and lattice vibrations, observing higher-order phonon modes extending up to the fourth order. We observe an anomalously high and weakly temperature-dependent intensity ratio of these higher-order modes relative to low-energy first-order phonons. Theoretical cluster calculations and Franck-Condon modelling reveal that these features originate from resonantly enhanced multiphonon scattering mediated by localized intermediate Ni $3d^8$ multiplet excitations. The local orbital occupancy of these intermediate states couples strongly to lattice coordinates, providing a selective enhancement mechanism for specific phonon channels. Notably, these higher-order features are absent in the end-member $Fe_2P_2S_6$. This distinction reflects the charge-transfer character of $Ni_2P_2S_6$, which places optical transitions in resonance with local multiplet states, contrasted with the Mott-Hubbard insulator regime in Fe-rich analogues. Our findings clarify the microscopic origin of high-frequency Raman scattering in transition metal thiophosphates and underscore the central role of local orbital-phonon interactions in 2D correlated magnets.

*nasarukhan736@gmail.com
†pkumar@iitmandi.ac.in

## 1. Introduction

Van der Waals (vdW) magnetic materials have attracted considerable attention due to their novel physical properties and promising applications in spintronics and quantum electronics [1-7]. Among them, transition metal phosphorus trisulfides ($TM_2P_2S_6$) represent a distinct class of layered materials where adjacent layers interact through weak vdW forces, resulting in a quasi-two-dimensional magnetic structure. Recently, nickel phosphorus trisulfide ($Ni_2P_2S_6$) has emerged as a system of particular interest due to its rich physics, including phonon-magnon coupling, strong electronic correlations, and spin-correlated excitonic phenomena [8-12]. In $Ni_2P_2S_6$, antiferromagnetic zigzag order sets in below $T_N$ ~ 153 K [13]. Furthermore, $Ni_2P_2S_6$ can be exfoliated down to few-layer thicknesses, providing an ideal playground to examine quasiparticle dynamics across dimensions.

In correlated 3*d* transition metal compounds, the orbital degrees of freedom (DoF) often play a decisive role alongside spin and charge [14-21]. In isolated transition metal ions, 3*d* orbitals are degenerate, but local crystal fields break this degeneracy. In $Ni_2P_2S_6$, the octahedral coordination splits the Ni $3d^8$ shell into a fully occupied lower $t_{2g}$ triplet and a half-filled higher $e_g$ doublet, establishing a high-spin S = 1 state, as illustrated in Figure 1 [22]. Collective orbital excitations (orbitons) can emerge either in systems with long-range ordering of active, nearly degenerate orbitals such as the ($e_g^1$) manganites or ($t_{2g}^1$) titanates [23-24] or, at substantially higher energies, in systems where the orbital degeneracy is lifted by the local crystal field, as in the ($d^9$) cuprates [25-26]. In the latter case, the excitations originate from crystal-field-split multiplets but may nevertheless propagate coherently through intersite exchange and hybridization, and are often referred to as orbitons or ligand-extended excitons. Consequently, in $Ni_2P_2S_6$, such excitations are expected only at relatively high energies [8-12,22]. The low-energy orbital physics should therefore arise only indirectly, for example through orbital-phonon coupling, and is not expected to exhibit coherent collective orbital dynamics.

Raman spectroscopy is a powerful probe for quasiparticle excitations and coupling mechanisms [8, 27-38]. While previous Raman studies on 2D thiophosphates focused predominantly on first-order zone-center phonons and magneto-optical responses, high-frequency features lying beyond the first-order phonon manifold remain less understood.

In this work, we present a systematic temperature-dependent Raman study of bulk single crystals and thin flakes of $Ni_2P_2S_6$, as well as the series of $(Ni_xFe_{1-x})_2P_2S_6$. We report prominent, higher-order Raman features extending up to ~ 1600 $cm^{-1}$. Through a combination of thickness-dependent measurements, temperature tracking, and cluster-model calculations, we demonstrate that these high-frequency features correspond to higher-order multiphonon processes. These processes are strongly enhanced by resonant optical transitions through intermediate local Ni $3d^8$ multiplet states that couple efficiently to lattice vibrations.

## 2. Experimental details

Unpolarized Raman spectroscopic measurements were performed on single crystals and thin layers of $Ni_2P_2S_6$ as well as bulk single crystals of $(Ni_xFe_{1-x})_2P_2S_6$ (x= 0.7, 0.5, 0.3, 0) series, using a LabRam HR evolution micro-Raman spectrometer in the backscattering geometry. For $Ni_2P_2S_6$ (both bulk and exfoliated flakes), a 633 nm laser combined with a long working distance 50× objective was used. For $(Ni_xFe_{1-x})_2P_2S_6$ (x= 1, 0.7, 0.5, 0.3, 0) series, a 532 nm laser was used. The laser power at the sample was kept very low (~ 1 mW) to minimize the heating effects. All the measurements were carried out across a temperature range of 4 K to 300 K, with a temperature accuracy of ± 1K. Thin flakes of $Ni_2P_2S_6$ were prepared via the standard mechanical exfoliation method on $SiO_2$/Si substrates. The flake thicknesses were subsequently determined using Atomic Force Microscopy (AFM).

## 3. Results and discussion

### 3.1 First-order Raman scattering

$Ni_2P_2S_6$, in its bulk and few-layer forms (excluding the monolayer), crystallizes in a monoclinic structure and belongs to the point group $C_{2h}$ and space group $C2/m$. The Ni atoms form a hexagonal arrangement, each surrounded by six S atoms in a trigonal coordination. Two P atoms bond covalently with six S atoms, forming the anionic complex $(P_2S_6)^{-4}$ (see Supplementary Figure S1) [39]. In both bulk and few-layer systems, the unit cell of $Ni_2P_2S_6$ gives rise to 30 phonon modes at the Γ-point of the Brillouin zone, with irreducible representations $8A_g + 6A_u + 7B_g + 9B_u$. Among these, 15 modes are Raman-active ($\Gamma_{Raman} = 8A_g + 7B_g$), 12 are infrared-active ($\Gamma_{IR} = 5A_u + 7B_u$), and 3 correspond to acoustic phonons ($\Gamma_{Acoustic} = A_u + 2B_u$) [40]. Figure 2 presents the raw Raman spectra collected at 4 K for flakes of $Ni_2P_2S_6$ with thicknesses of 40 nm (treated as bulk), 20 nm, 15 nm, and 10 nm, covering a spectral range of 50 $cm^{-1}$ to 1730 $cm^{-1}$ (see Table S1 for mode frequency values at 4 K).

For clarity, we have labelled the observed phonon modes as P1-P22 and S1-S3. The assignment includes 10 first-order phonon modes for bulk, 12 for the 20 nm flake, and 13 for the 15 nm and 10 nm flakes of $Ni_2P_2S_6$. These first-order phonon modes in $Ni_2P_2S_6$ may be categorized into external modes, which falls below ~ 250 $cm^{-1}$ and are associated with the vibrations of heavy metal ions; and internal modes, which occur covers the spectral range of ~ 250 to ~ 600 $cm^{-1}$ and are linked to the vibrations of the anion complex [8, 41-44]. Our main focus in this paper is on the high frequency modes (i.e., above 650 $cm^{-1}$), which lie beyond the range of first-order phonon modes. To the best of our knowledge, these higher energy modes have not been explored in detail in earlier studies.

**3.2 Higher order Raman scattering**

The Raman spectra of both bulk and few-layered $Ni_2P_2S_6$ samples reveal several distinct features which are beyond the range of first-order phonons. As illustrated in Figure 3, the spectra recorded in the 700 - 1730 $cm^{-1}$ range showcase higher-order Raman-active modes. These spectral features are deconvoluted using a sum of Lorentzian functions to extract self-energy parameters i.e. mode frequencies, full width at half maximum (FWHM)/linewidths, as well as the intensities of the individual modes. Two prominent clusters dominate the high frequency Raman spectra, centered around ~ 800 $cm^{-1}$ and ~ 1170 $cm^{-1}$. A broad and very weak feature is also observed near 1600 $cm^{-1}$, which is fitted with a Gaussian line shape.

For bulk and exfoliated flakes (20 nm, 15 nm, and 10 nm), the spectra centred around ~ 800 $cm^{-1}$ can be well described by fitting 8 Lorentzian peaks (labelled as P11-P18). These higher-energy modes centred around ~ 800 $cm^{-1}$ may be attributed to the overtone and combination of the intense first-order P6 mode (~ 385 $cm^{-1}$). Doubling the frequency of the P6 mode yields a value close to the second-order modes centred at ~ 756 $cm^{-1}$ (P12), ~ 774 $cm^{-1}$ (P13), and ~ 788 $cm^{-1}$ (P14), consistent across both bulk and thin flake samples (see Supplementary Fig. S3 for comparing the spectra) [39]. While the proximity of these high-frequency modes (P11-P18) to the doubled of frequencies of first-order modes suggests their origin in second-order phonon scattering. However, several distinct characteristics of these modes (P11-P18) are not consistent with normal second-order phonon contribution picture: (i) no confined cluster of first-order phonon modes exist below 650 cm$^{-1}$, therefore a cluster of second-order phonon is generally not expected. (ii) Intensities of the higher-order peaks is anomalously high, even modes up to fourth-order are observed and (iii) these modes are nearly evenly spaced separated by ~ 20 cm$^{-1}$. (iv) Intensity ratio of these modes w.r.t their first-order counterpart is anomalously high and nearly temperature independent. Taken together, these observations suggest the role of resonance phenomena involving localized Ni $3d^8$ multiplets and local orbital-phonon coupling, discussed in detailed in section 3.5 and the assignment of these modes in Table 1.

In the spectral region above 1000 cm$^{-1}$, the Raman response exhibits a well-defined cluster of modes centered near ~ 1170 cm$^{-1}$ (see Fig. 3). This cluster is resolved into three distinct Lorentzian peaks. Notably, modes P20 and P21, centered at ~ 1176 cm$^{-1}$ and ~ 1192 cm$^{-1}$ respectively, may be attributed to second-order overtone contributions of the P9 mode ( ~ 591 cm$^{-1}$), as the doubled of frequency of P9 lies close to these peak positions ( $2\omega_{P9} = 1182$ cm$^{-1}$). Interestingly, for the bulk, 20 nm, and 15 nm samples, both the ~ 800 cm$^{-1}$ and ~ 1170 cm$^{-1}$ clusters persist up to room temperature (300 K), though the intensity of the side bands becomes very weak. However, in the 10 nm sample, all modes in these clusters - except for the most intense modes i.e. P16 and P21 disappear above ~ 225 K (see Supplementary Fig. S2) [39], even intensity of these modes, i.e. P16 and P21, becomes too weak. Much like the ~ 800 cm$^{-1}$ cluster, the behaviour of the ~ 1170 cm$^{-1}$ cluster modes deviate from expectations for normal higher-order phonon modes; suggesting the role of resonance phenomena and local orbital-phonon coupling.

Figure 4 presents the temperature dependence of the peak frequency and FWHM for modes P11, P15, P16, and P21 in bulk, 20 nm, 15 nm, and 10 nm $Ni_2P_2S_6$ samples. Several key observations are: (i) all modes exhibit typical phonon softening with increasing temperature. However, for the 15 nm flake, the P15 mode shows an anomalous hump near 150 K. A notable change in slope is observed around ~150 K for all the modes in the bulk sample: in particular for modes P11, P15, and P16 in the 20 nm flake and modes P15 and P16 in the 15 nm flake and modes P15 and P16 in the 10 nm flake. (ii) FWHM of P16 and P21 modes display conventional broadening with rising temperature across all thicknesses, including bulk. However, a change in slope is again detected near ~150 K. (iii) FWHM of P15 mode decreases with decreasing temperature, but below ~ 150 K, it exhibits a reverse trend i.e. broadening as the temperature drops further. (iv) For bulk and 10 nm samples, FWHM of P11 mode increases with increasing temperature, while in 20 nm and 15 nm flakes, it remains nearly temperature-independent. It is noteworthy that the antiferromagnetic transition in $Ni_2P_2S_6$ occurs at ~ 153 K [8, 45-46]. The anomalies in both peak position and FWHM near this temperature suggest a coupling between vibrational and magnetic degrees of freedom, especially in proximity to the paramagnetic-to-antiferromagnetic phase transition in both bulk and few-layer samples.

### 3.3 Resonance phenomena and local orbital-phonon coupling in $Ni_2P_2S_6$

By analysing the Raman-active modes within the 700-1200 $cm^{-1}$ spectral range, one might initially interpret them as normal second-order phonon modes, given that some of their frequencies approximately correspond to twice those of specific first-order phonon modes. However, a detailed analysis of their temperature-dependent intensity reveals contributions from resonance phenomena and local orbital-phonon coupling, discussed in detailed in section 3.5; we note that the role of orbital degrees of freedom and their coupling with lattice vibrations have been extensively investigated in *3d* transition metal oxides [14,19-21, 47-52].

Assigning second-order Raman bands corresponding to specific first-order phonon modes is inherently challenging, as second-order scattering involves phonons across the entire Brillouin zone (BZ). Notably, regions with a higher phonon density of states contribute more significantly, resulting in broader second-order features compared to their first-order counterparts. Additionally, the peak frequencies of these second-order modes do not necessarily equate to twice the frequency of the corresponding first-order modes at the Γ-point. The Raman mode centred at ~ 818 $cm^{-1}$ (P12, P13) may be considered as a second-order counterpart of the first-order P6 mode (~ 385 $cm^{-1}$). Similarly, the mode at ~ 1192 $cm^{-1}$ (P21) and the very weak feature near ~ 1600 $cm^{-1}$ (P22) could represent the third- and fourth-order overtones of the P6 mode, respectively. Figure 5 presents the temperature evolution of the intensity ratios i.e. second-order to first-order ($I_{16}/I_6$, and $I_{21}/I_9$) and third-order to first-order ($I_{21}/I_6$). The fourth-order to first-order ratio ($I_{22}/I_6$) is shown only for the bulk system (see Fig. 7 (a)). Generally, second- and higher-order Raman modes are expected to exhibit much weaker intensities compared to their first-order counterparts [53]. To test this, we plotted the intensity ratios $I_{16}/I_6$ (P16/P6), $I_{21}/I_6$, and $I_{22}/I_6$ for both thin flakes and bulk $Ni_2P_2S_6$. Quite unexpectedly, the second- and third-order modes exhibit intensities comparable to or even exceeding that of the first-order P6 mode. Although the fourth-order mode remains weaker, its intensity ratio (~ 0.15 - 0.4) is still substantially high. These anomalously strong higher-order intensities challenge their straightforward assignment as conventional multiphonon overtones or combination, suggesting the role of resonance phenomena and local orbital-phonon coupling. The assignment of these modes as overtones, combination of first-order modes and is listed in Table 1 and discussed in section 3.5.

Typically, higher-order Raman modes are expected to be both broader and significantly weaker than their corresponding first-order counterparts. According to perturbative considerations, the intensity of an *n*th-order phonon mode scales as $g^n$, where $g$ is the electron–phonon coupling constant (with $g < 1$). Consequently, the intensity should diminish rapidly with increasing scattering order. However, our measurements reveal an intriguing deviation from this expectation: the intensity ratios between higher-order and first-order modes remain nearly constant-or increases slightly with decreasing temperature. The observed enhancement of the intensity ratios at lower temperatures may be understood within the framework of resonant Raman scattering. In this picture, the enhanced intensity at lower temperatures may be ascribed to a longer lifetime of the resonant excitation [54]. More specifically, the Raman intensity ratio between second- and first-order modes is proportional to the square of the excited-state lifetime at resonance.

Notably, a similar temperature-independent intensity behaviour was advocated by Allen *et al.* [48] invoking orbital DoF. These anomalous temperature-dependent trends observed here suggest that these high-frequency Raman features are not normal multiphonon rather results from the resonant enhanced multiphonon scattering and localized orbital-phonon coupling. To understand this observed behaviour, we explore two established theoretical models for orbiton-phonon dynamics in *3d* systems. (i) Brink's model [55] treats electron-electron correlations and lattice vibrations on equal footing. It shows that orbiton dispersion which is primarily driven by electron correlation is significantly renormalized by strong electron-phonon coupling. The resulting interaction yields satellite features in the Raman spectrum and gives rise to hybridized orbiton-phonon excitations at higher energy scales. We note that this model is unlikely to be active in this system, discussed in section 3.5.2(B) (ii) Allen's model [48] considered local rearrangement of atoms in the octahedral coordination, which leads to the formation of self-trapped excitons. Within this context, a Franck-Condon process generates unusually intense, Raman-active multiphonon bands, having almost same intensity as their first-order counterpart. In our case, very high intensity of the multiphonon bands as well as the emergence of multiple satellite modes suggest strong contributions from the localized orbitals and the resonance phenomena.

We also examined an alternative scenario wherein the high-energy Raman mode P21 may be interpreted as a normal second-order overtone of the P9 mode, given that its frequency is approximately double that of P9. To evaluate this, we plotted the intensity ratio $I_{21}/I_9$ (see bottom panel of Fig. 5). Interestingly, the ratio is significantly high (~ 2-3) and remains nearly constant across the entire temperature range. This anomalously enhanced intensity behaviour deviates from the conventional multiphonon processes, suggesting resonantly enhanced multi-phonon scattering involving localized Ni $3d^8$ multiplets.

### 3.4 Resonance phenomena and local orbital-phonon coupling in $(Ni_xFe_{1-x})_2P_2S_6$ series

We further performed Raman scattering measurements on a series of single crystals with varying Fe substitution in $(Ni_xFe_{1-x})_2P_2S_6$, where $x$ = 1.0, 0.7, 0.5, 0.3, and 0. Supplementary Figure S5 [39] shows the Raman spectra of this compositional series, and supplementary section A provides a detailed description of the first-order phonon modes and two-phonon scattering processes for the bulk $(Ni_xFe_{1-x})_2P_2S_6$ samples. Notably, the characteristic Raman modes P6, P9, P16, P21, and P22 observed in $Ni_2P_2S_6$ also appear in the $x$ = 0.7, 0.5, and 0.3 compositions. However, these high energy clusters centred around ~ 800 $cm^{-1}$ and ~ 1170 $cm^{-1}$ are conspicuously absent in the end-member $Fe_2P_2S_6$.

To further investigate this behaviour, we qualitatively analysed the intensity ratios of higher-order to first-order modes in the doped compositions. Figures 6 and 7 display the ratios: $I_{16}/I_6$ (P16 to P6), $I_{21}/I_6$ (P21 to P6), $I_{21}/I_9$ (P21 to P9), and $I_{22}/I_6$ (P22 to P6) for the $x$ = 0.7, 0.5, and 0.3 samples. Even in the doped systems, the higher-order modes exhibit intensities that are comparable to or even exceed their first-order counterparts. Additionally, many evenly spaced modes are observed around the prominent features near ~800 $cm^{-1}$ and ~1170 $cm^{-1}$. These anomalously high intensities of these high energy modes suggest resonantly enhanced higher order multiphonon scattering and local orbital-phonon coupling involving Ni $3d^8$ multiplets.

It is both intriguing and unexpected that anomalous higher order Raman scattering signatures are absent in $Fe_2P_2S_6$. We have invoked electronic classification i.e. charge transfer and Mott-Hubard insulator nature of these systems to understand the absence of higher order phonon modes in $Fe_2P_2S_6$. In $Ni_2P_2S_6$, the system is classified as a charge-transfer insulator, where the energy required for transferring an electron from the S *3p* orbital to the Ni *3d* orbital is lower than the on-site Coulomb repulsion within the Ni *3d* orbitals. This energetic configuration allows for appreciable ligand-to-metal charge transfer, modifying the occupancy of the Ni *3d* orbitals. Such selective orbital population can lower the total energy and render the system susceptible to Jahn-Teller distortions, thereby favouring active role of local orbital. In contrast, $Fe_2P_2S_6$ is classified as a Mott-Hubbard insulator, where the charge transfer energy between Fe *3d* and S *3p* orbitals is greater than the on-site Coulomb repulsion of the Fe *3d* electrons. Consequently, ligand-to-metal charge transfer is energetically unfavourable, resulting in relatively rigid orbital occupancy and loss of optical resonance. This rigidity hinders both Jahn-Teller distortions and the emergence of local orbital-phonon coupling in $Fe_2P_2S_6$.

The substitution of Fe into $Ni_2P_2S_6$ provides a unique tuning knob to explore the interplay between electronic correlations, lattice dynamics, spin and orbital DoF across the $(Ni_xFe_{1-x})_2P_2S_6$ system. Our Raman measurements across the doping series ($x$ = 1.0 to 0.0) reveal a remarkable evolution of the intense higher-order Raman modes associated with resonantly enhanced multiphonon scattering and local orbital-phonon coupling suggesting a direct correlation with changes in the underlying spin, orbital and electronic character. We note that we observed strong first-order phonon modes (P8-P10) centred around ~ 600 $cm^{-1}$, only for pure Ni or Ni doped samples (see Fig. S5), suggesting a robust coupling between phonons associated with the $P_2S_6$ cage and the charge transfer between Ni and Sulphur atoms or a resonance may be happening with the process of charge transfer and this resonance may give rise to these strong phonon modes as well as high energy phonon modes. However, these phonon modes near ~ 600 $cm^{-1}$ in case of $Fe_2P_2S_6$ are absent suggesting minimum or no coupling between phonons and charge transfer process, or no charge transfer is possible hence no resonance and as a result no observation of these first-order as well as high energy phonon modes [8].

The intermediate compositions (from $x$ = 0.7 to 0.3) show a progressive weakening of the higher order modes with increasing Fe content, suggesting a dilution of orbital fluctuations in tandem with a transition from charge-transfer to Mott-Hubbard behaviour. Still, the persistence of resonance phenomena and local orbital-phonon coupling down to $x$ = 0.3 underscores the robustness of the orbital DoF in influencing lattice dynamics. Our experimental results reinforce the role of orbital DoF in Ni-rich compounds, positioning them as candidates for exploring spin-orbital entanglement in these 2D magnetic systems. We hope that our experimental results will motivate further research in this direction on these systems to understand the role of resonance phenomena and orbital degrees of freedom.

### 3.5 Theoretical calculations and discussion

#### 3.5.1 Interpretation of the $Ni_2P_2S_6$ Raman spectrum: Two-phonon peaks scenario

#### A. Preliminaries

Resonant Raman scattering process occurs when the excitation energy lies close to an allowed optical transition of the material. In that case the Raman process is enhanced because it proceeds through real (or nearly real) intermediate electronic states. This can be made explicit by writing the Raman amplitude schematically as the resonant part of the Kramers-Heisenberg-Dirac expression as [56-57]

$$A_{Raman} \propto \sum_m \frac{\langle f | D | m \rangle \langle m | D | i \rangle}{E_i + \hbar\omega_{exc} - E_m + i\Gamma_m} \tag{1}$$

where $D$ is the effective dipole operator, $|m\rangle$ denotes optically excited intermediate states with energy $E_m$ and life time broadening $\Gamma_m$. Finally, $|i\rangle, |m\rangle, |f\rangle$ denotes ground, intermediate, and final states of the Raman process that are eigenstates of the electron-phonon Hamiltonian $H$ with energies $E_i, E_m, E_f$; respectively. In the resonant limit the incoming photon frequency $\hbar\omega_{exc} \approx E_m - E_i$, the detuning $\Delta = E_m - (E_i + \hbar\omega_{exc})$ becomes small, and the relatively small denominator in Eq. (1) strongly enhances $A_{Raman}(\omega_{exc})$ [56-57].

A simple microscopic way to understand the appearance of multiphonon Raman features is provided by the displaced-harmonic-oscillator, or Franck-Condon, picture. If the resonant intermediate electronic state has an equilibrium phonon coordinate shifted by ($\Delta Q$) with respect to the ground state, the optical transition connects the vibrational ground state to a ladder of vibrational states. The corresponding Franck-Condon weights are controlled by the Huang-Rhys factor

$$S = \frac{M\Omega}{2\hbar}(\Delta Q)^2 \tag{2}$$

where ($M$) is the effective mass of the vibrational mode, ($\Omega$) is its angular frequency, and ($\Delta Q$) is the displacement between the equilibrium coordinates of the ground and intermediate electronic states. In the simplest one-mode model, the probability of reaching a final state with ($n$) phonons is then

$$P_n = e^{-S}\frac{S^n}{n!} \tag{3}$$

Thus, for sizeable ( $S$ ), the Raman intensity can be distributed over several multiphonon final states rather than being restricted to the one-phonon channel. Importantly, resonance does not by itself create these channels; rather, through the small Kramers-Heisenberg-Dirac energy denominator, it strongly enhances the scattering amplitude through intermediate vibronic states for which the Franck-Condon overlaps to multiphonon final states are non-negligible [48,58-60].

**B. Peak assignment**

In the current Raman experiment, the incoming photons have wavelength $\lambda_{exc}$ = 633 nm, corresponding to the energy $\hbar\omega_{exc}$ ~ 1.96 eV. Figure 8(a) compares the experimental optical absorption spectrum of bulk $Ni_2P_2S_6$ at low temperature with the theoretical multiplet spectrum obtained for the $NiS_6$ cluster with the Ni $3d^8$ configuration (For detail of the model and its parameters see [22, 61]). The calculation, shown in more detail in Fig. 8 (b c), reveals local multiplet excitations in the range 1.5-2.0 eV (labelled $|E4\rangle$ and $|E5\rangle$). While the absorption rises rapidly above ~ 2.0 eV, consistent with the onset of charge-transfer excitations, the lower-energy peaks in the optical absorption spectrum correspond primarily to local multiplet excitations. Since $\hbar\omega_{exc}$ ~ 1.96 eV lies close to intrinsic optical transitions in $Ni_2P_2S_6$ that predominantly involve these multiplet excitations, we believe that the observed Raman spectrum is resonantly enhanced via intermediate state(s) of predominantly multiplet character.

The resonant picture provides a natural organizing principle for the Raman spectra summarized in Table I. The most intense first-order (i.e. 1R) modes are the prominent phonons near ~ 385 $cm^{-1}$ (P6) and ~ 590-595 $cm^{-1}$ (P9/P10), which strongly modulate the polarizability. Under resonant conditions, overtones and combinations involving these modes become visible and can dominate the spectrum. In particular, the dense set of peaks observed between 700 - 900

cm$^{-1}$ is consistent with P6-based two-phonon processes (overtones and Brillouin-zone combinations), while the features near 1170-1200 cm$^{-1}$ are naturally associated with overtones/combinations of the high-energy P9/P10 manifold. Thus, the proximity of 633 nm excitation to optical transitions (Fig. 8) explains why multiphonon scattering is strongly enhanced and why the structure summarized in Table I is expected in the resonant regime.

**C. Role of orbital degrees of freedom**

The resonant character of the Raman scattering in $Ni_2P_2S_6$ also allows the observed spectral features to be discussed in terms of orbital-phonon coupling. As discussed above, the excitation energy is resonant with primarily intra-atomic multiplet transitions of the $Ni^{2+}$ ion (see Fig. 8), so that the intermediate Raman state corresponds to a localized electronic excitation with a modified orbital occupancy. In other words, the Raman process transiently creates a local orbital excitation that differs from the ground state in its $e_g/t_{2g}$ character and ligand-hole admixture. Because the energy of such a multiplet state depends sensitively on the local crystal-field environment, it is naturally coupled to lattice distortions. This coupling can be expressed phenomenologically by allowing the multiplet energy $E_m$ to depend on the phonon coordinate $Q$,

$$E_m(Q) = E_m^{(0)} + gQ + o(Q^2)\,, \tag{4}$$

where $g = \partial E_m / \partial Q$ is an effective orbital-phonon coupling constant. A nonzero $g$ implies that lattice vibrations modulate the energy of the orbital excitation, providing a microscopic mechanism for strong phonon involvement in the resonant Raman process. Within the Raman scattering amplitude, this dependence enters through the resonance denominator,

$$\frac{1}{E_i + \hbar\omega_{exc} - E_m(Q) + i\Gamma_m} \tag{5}$$

so that fluctuations of $Q$ strongly affect the scattering probability when the excitation is close to resonance. As a result, phonon modes that couple efficiently to the local orbital excitation are selectively enhanced. This naturally explains why a small subset of first-order phonons (most notably the modes near ~ 385 $cm^{-1}$ and ~ 590-595 $cm^{-1}$) dominate both the single-phonon Raman response and the higher-order overtone and combination bands summarized in Table I. Note that the orbital excitations involved here are localized multiplet states rather than propagating collective orbitons [14, 25-26, 62]. In this sense, the present experiment probes the interaction between local orbital degrees of freedom and phonons, revealed through resonantly enhanced multi-phonon and double-resonant Raman scattering.

### 3.5.2 Scenario that can be excluded

Below we discuss a simple electronic scenarios that might a *priori* be capable of explaining the observed Raman spectrum, yet can ultimately be excluded. Note also that perhaps the simplest electronic scenario, namely intersite charge excitations, can only occur at energies above the charge gap of $Ni_2P_2S_6$, which is safely above 1 eV [63], and therefore cannot account for the observed low-energy Raman features.

### A. Local multiplet scenario

The simplest interpretation of the observed Raman peaks would be to associate them with local multiplet transitions. From the above considerations, see Fig. 8, it is clear that all but three "multiplet" eigenstates of the $NiS_6$ cluster lie above ~ 0.8 eV and therefore cannot account for the Raman peaks under consideration. Hence, the only possible candidates are the high-spin "ground state" multiplets. These states are degenerate at zero magnetic field, but their energies split linearly with increasing applied field acting on the cluster. To investigate the above possibility, we repeated the $NiS_6$ cluster calculations presented above (see Fig. 8), now including an external magnetic field in order to mimic the effect of magnetic ordering. Indeed, when the external field reaches approximately 50 meV, the high-spin multiplets with different $S_Z$ quantum numbers ($S_Z$=0,1) acquire excitation energies close to the experimentally observed Raman peaks. Nevertheless, we consider such an interpretation unlikely for the following reasons. First, the magnitude of the external field required to reproduce these excitation energies is unrealistically large compared to the spin exchange energy scales present in the system [64]. Second, in a realistic crystal these local magnetic excitations are not strictly localized but become dispersive, which substantially lowers their energy at small momenta. In fact, they evolve into magnon ($S_Z = 0$) and two-magnon ($S_Z = 0$) excitations, which are indeed observed in $Ni_2P_2S_6$ at approximately 0.03 and 0.08 eV, respectively [22].

### B. Orbitons in Raman due to the Jahn-Teller effect

A Jahn-Teller-type mechanism of the kind proposed in Ref. [55] is unlikely to operate in the present case. The key limitation is the absence of sufficient orbital degrees of freedom at low energies. Indeed, the relevant low-energy excitations are restricted to the high-spin $e_g$ subspace, whereas both the high-spin states associated with Hund's coupling $J_H$ and the $t_{2g}$ excitations occur at substantially higher energies (of the order of 10Dq, i.e., well above ∼ 0.2 eV). As a result, the system effectively contains two high-spin electrons confined to the $e_g$ subshell. Under these conditions, a Jahn-Teller-like orbital splitting cannot occur, and consequently an "orbiton–phonon" scenario analogous to that discussed in Ref. [55] is not expected.

## Conclusion

Our detailed Raman study across the single crystals $(Ni_xFe_{1-x})_2P_2S_6$ series elucidates the origin of high-frequency Raman features in these 2D van der Waals magnets. In $Ni_2P_2S_6$, intense higher-order Raman modes extending up to ~ 1600 $cm^{-1}$ stem from resonantly enhanced multiphonon scattering. The excitation selectively targets local Ni *$3d^8$* intermediate multiplet transitions, where localized orbital occupancy couples strongly to lattice coordinates ( $g = \partial E_m / \partial Q$ ). This local orbital-phonon interaction amplifies overtone and combination channels via Franck-Condon and double-resonance mechanisms.

In contrast, the quenching of these higher-order modes in Mott-Hubbard $Fe_2P_2S_6$ reflects the loss of optical resonance and rigid orbital occupancy. Our findings establish $(Ni_xFe_{1-x})_2P_2S_6$ as a model platform to explore how electronic classification (charge-transfer vs. Mott-Hubbard) governs resonant optical interactions and local orbital-phonon coupling in low-dimensional materials.

**Acknowledgement**

NK acknowledge CSIR India for the fellowship. PK acknowledge support from IIT Mandi for the experimental facilities and SERB (CRG/2023/002069) for the financial support. S.A. acknowledges "MagTop" project (FENG.02.01-IP.05-0028/23) carried out within the "International Research Agendas" programme of the Foundation for Polish Science co-financed by the European Union under the European Funds for Smart Economy 2021-2027 (FENG). K.W. and M. R. acknowledge the support of National Science Centre in Poland under Project No. 2024/55/B/ST3/03144 and 2023/51/D/ST11/02588 (respectively). M.R. acknowledges access to the LEM computing cluster provided by the Wroclaw Centre for Networking and Supercomputing (WCSS).

**Figures:**

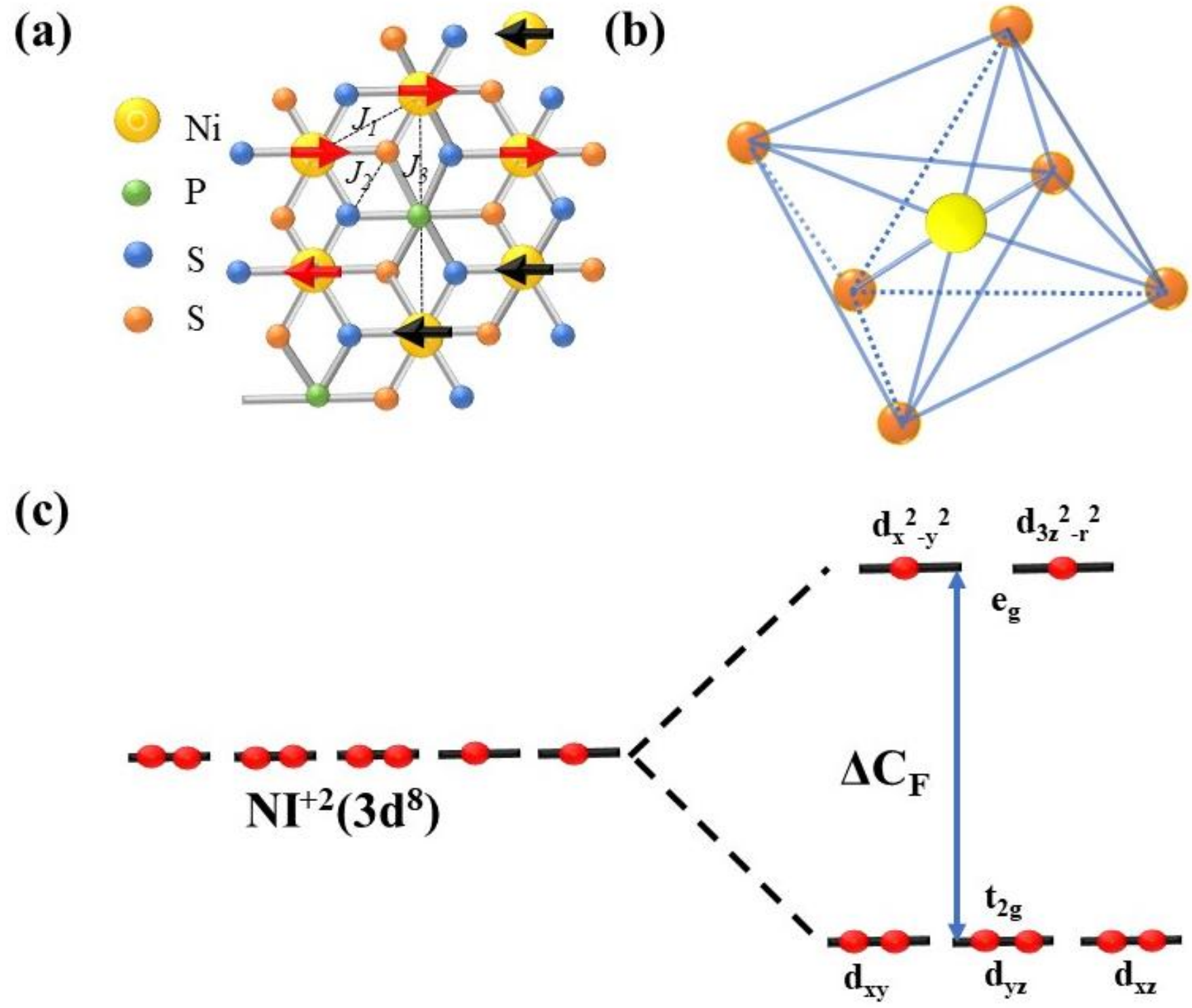

**Figure 1. (a)** Top-view illustration of the antiferromagnetic arrangement in $Ni_2P_2S_6$. Yellow, green, blue, and orange spheres represent Ni atoms, top-layer sulfur atoms, bottom-layer sulfur atoms, and phosphorus atoms, respectively. Red arrows on the Ni sites indicate spin orientations in the AFM zigzag magnetic phase. **(b)** Local octahedral crystal field environment surrounding a $Ni^{2+}$ ion in bulk $Ni_2P_2S_6$, highlighting the coordination with six sulfur ligands. **(c)** Schematic representation of $Ni^{2+}$ 3d orbitals: the left panel depicts the fivefold degenerate *d* orbitals in the absence of a crystal field, while the right panel shows their splitting into lower-energy $t_{2g}$ and higher-energy $e_g$ levels under an octahedral crystal field

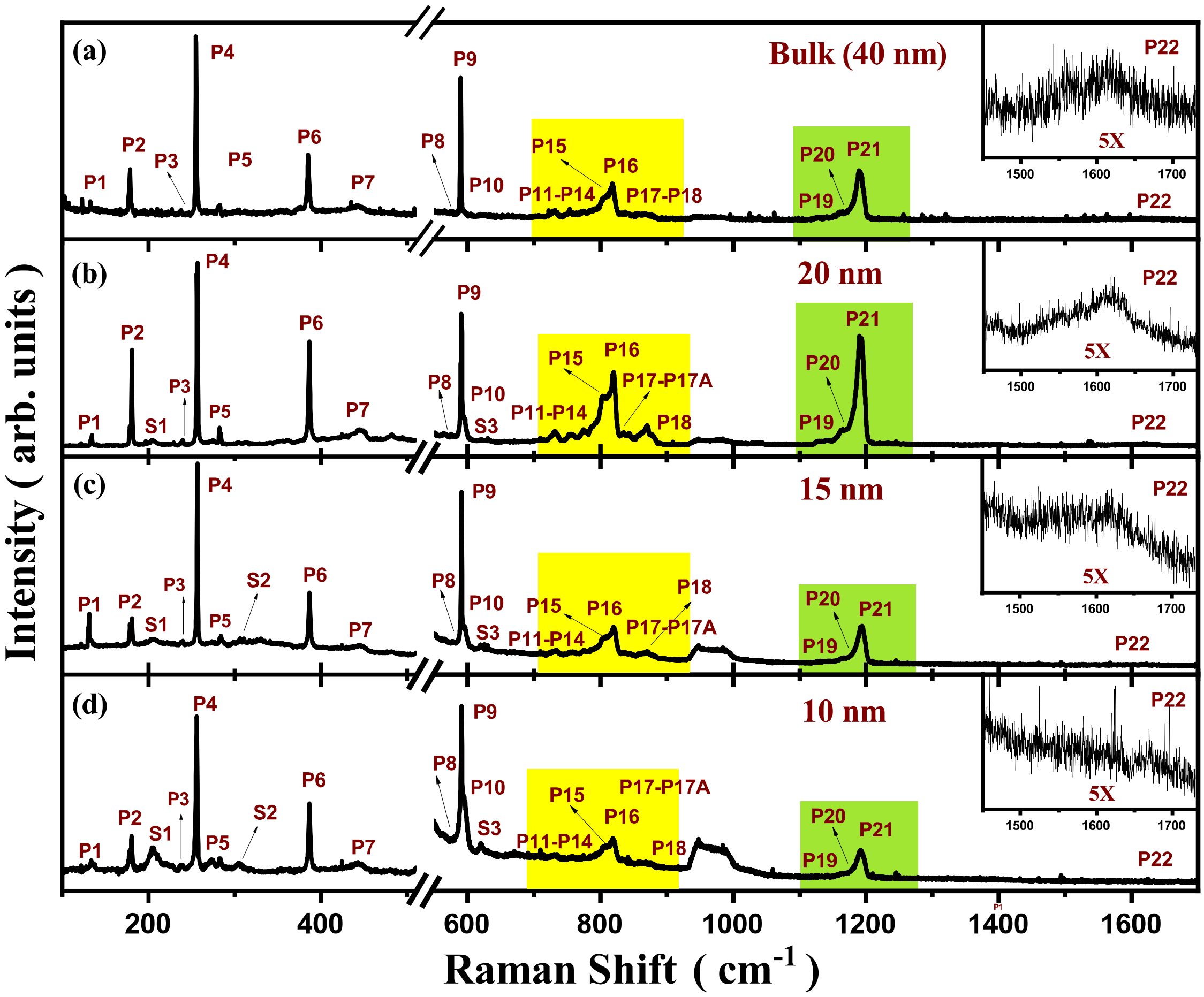

**Figure 2.** Raw Raman spectra measured at 4 K using a 633 nm excitation laser across the spectral range of 100-1700 cm$^{-1}$ for: **(a)** bulk $Ni_2P_2S_6$ (40 nm thick flake), **(b)** 20 nm $Ni_2P_2S_6$, **(c)** 15 nm $Ni_2P_2S_6$, and **(d)** 10 nm $Ni_2P_2S_6$. The yellow-shaded region highlights a cluster of peaks deconvoluted into eight Lorentzian components, while the green-shaded region denotes a cluster consisting of three Lorentzian peaks. Raman-active modes are labelled as P1-P22, including an additional split feature P17A, and secondary features labelled as S1-S3. Insets provide a magnified view of the weak high-frequency mode P22.

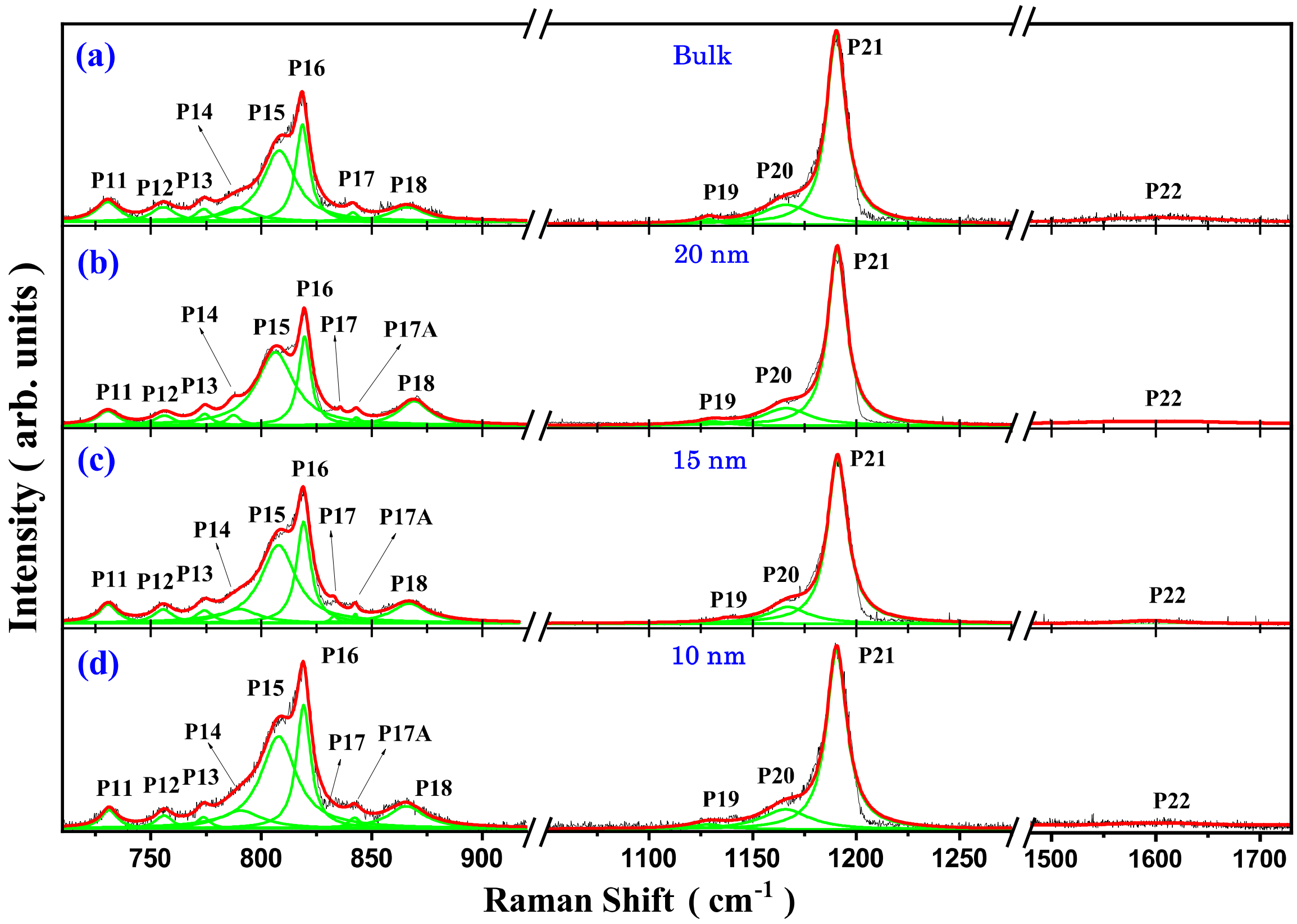

**Figure 3.** Fitted Raman spectra recorded at 4 K using a 633 nm excitation laser in the spectral range of 700-1730 cm$^{-1}$ for: **(a)** bulk $Ni_2P_2S_6$, **(b)** 20 nm $Ni_2P_2S_6$, **(c)** 15 nm $Ni_2P_2S_6$, and **(d)** 10 nm $Ni_2P_2S_6$. The solid red line represents the cumulative fit to the experimental data, while solid green lines denote the individual Lorentzian peak components used in the fitting process. Observed Raman features are labelled P11-P22, P17A.

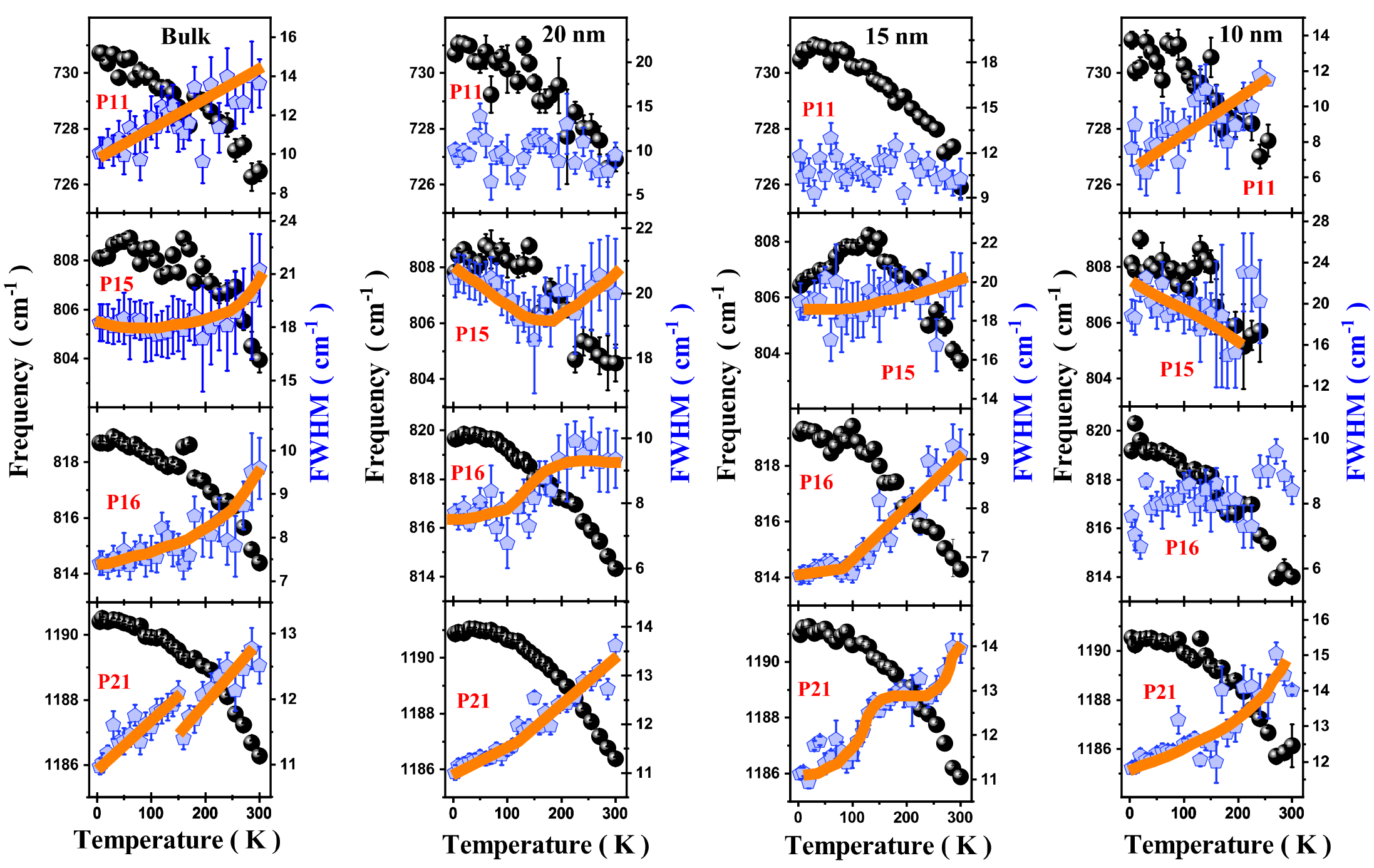

**Figure 4.** Temperature-dependent evolution of peak frequency and FWHM for Raman modes P11, P15, P16, and P21 across bulk $Ni_2P_2S_6$ and exfoliated flakes (20 nm, 15 nm, and 10 nm). Solid spheres represent the peak frequencies, while pentagons denote the corresponding FWHM values. Orange solid lines serve as visual guides for tracking the FWHM variation with temperature.

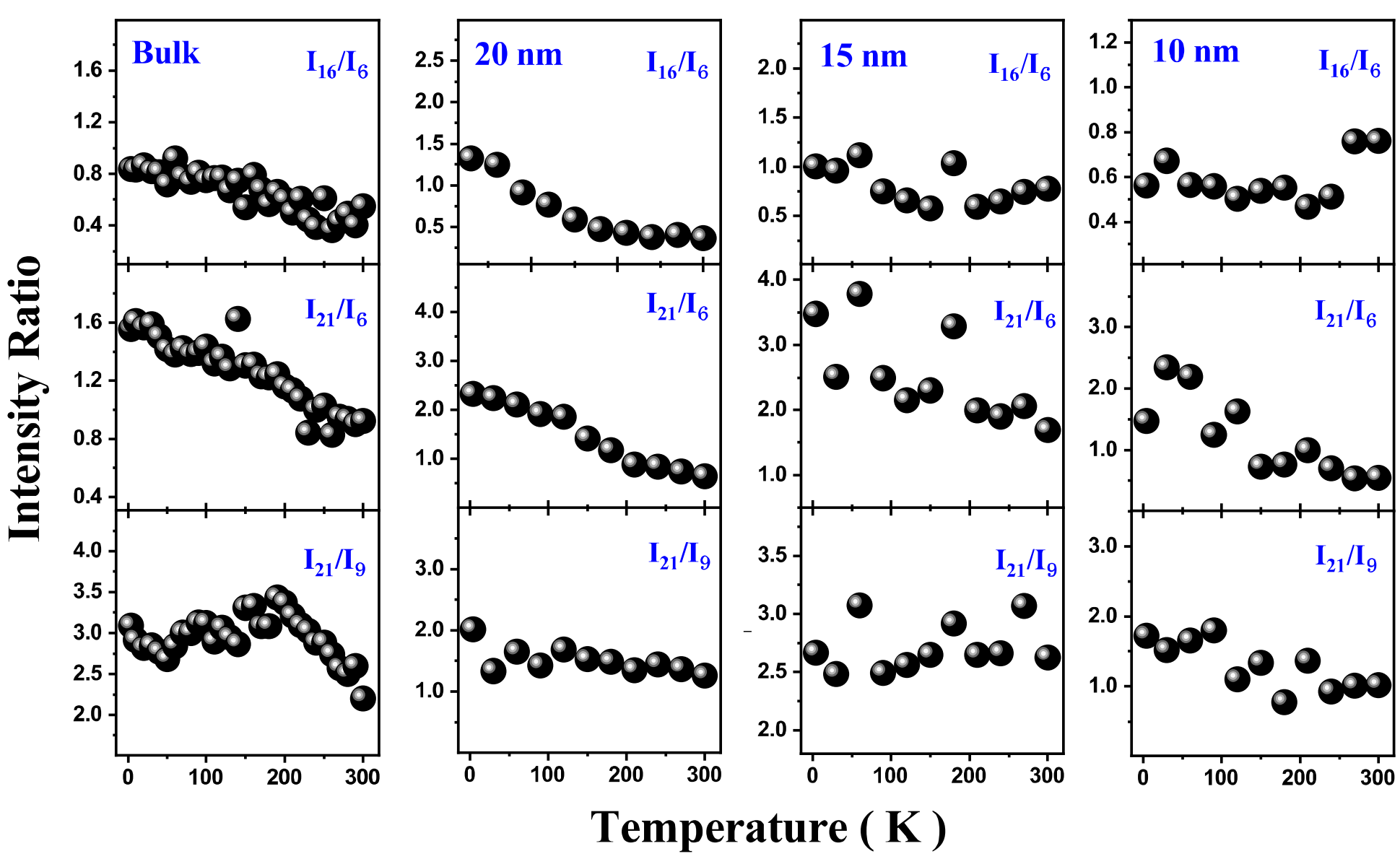

**Figure 5.** Variation of the intensity ratios $I_{16}/I_6$ (P16 to P6), $I_{21}/I_6$ (P21 to P6), and $I_{21}/I_9$ (P21 to P9) for *$Ni_2P_2S_6$* in (a) bulk, (b) 20 nm, (c) 15 nm, and (d) 10 nm thick flakes.

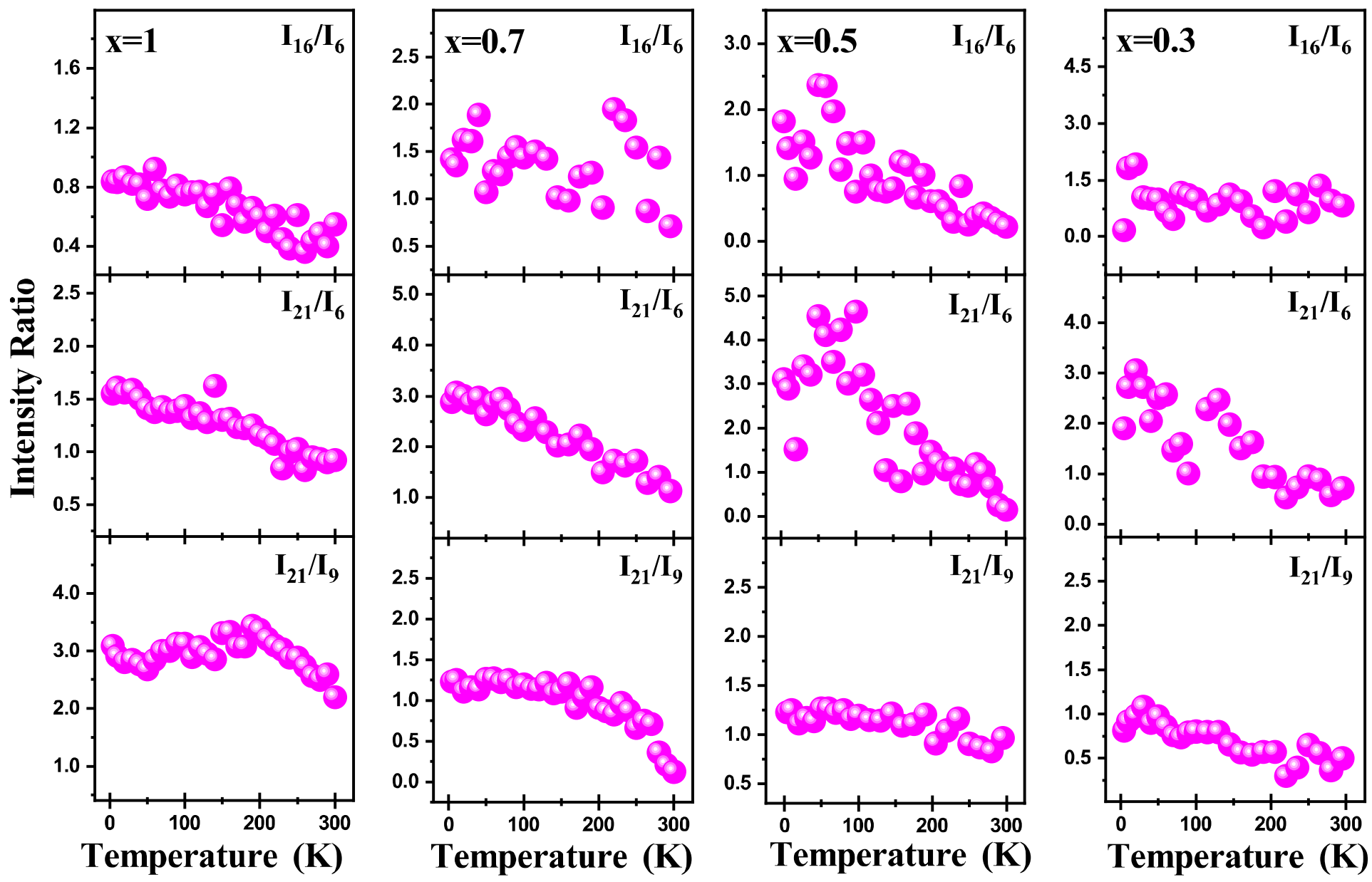

**Figure 6.** Comparison of intensity ratio $I_{16}/I_6$ ($P_{16}$ to $P_6$), $I_{21}/I_6$ ($P_{21}$ to $P_6$), and $I_{21}/I_9$ ($P_{21}$ to $P_9$) across the $(Ni_xFe_{1-x})_2P_2S_6$ series for compositions x = 1.0, 0.7, 0.5, and 0.3. These ratios highlight the evolution of higher-order mode intensities relative to their first-order counterparts, providing insights into orbiton-phonon coupling and its compositional dependence.

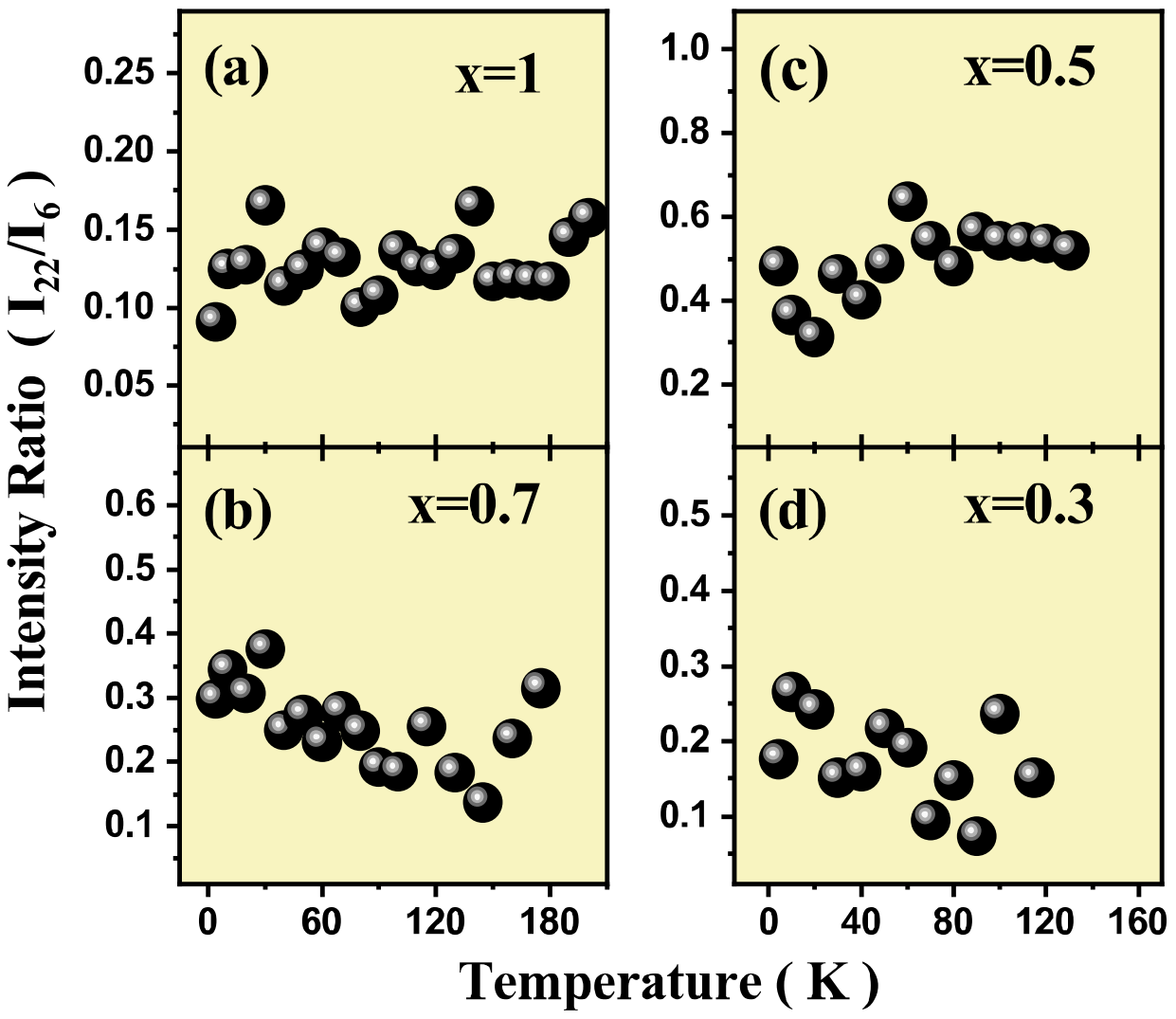

**Figure 7.** Intensity ratio of Raman mode P22 to P6 ($I_{22}/I_6$) for the $(Ni_xFe_{1-x})_2P_2S_6$ series with compositions: (a) $x$ = 1.0, (b) $x$ = 0.7, (c) $x$ = 0.5, and (d) $x$ = 0.3. This comparison highlights the variation in fourth-order Raman mode strength relative to its first-order counterpart across different Fe substitution levels.

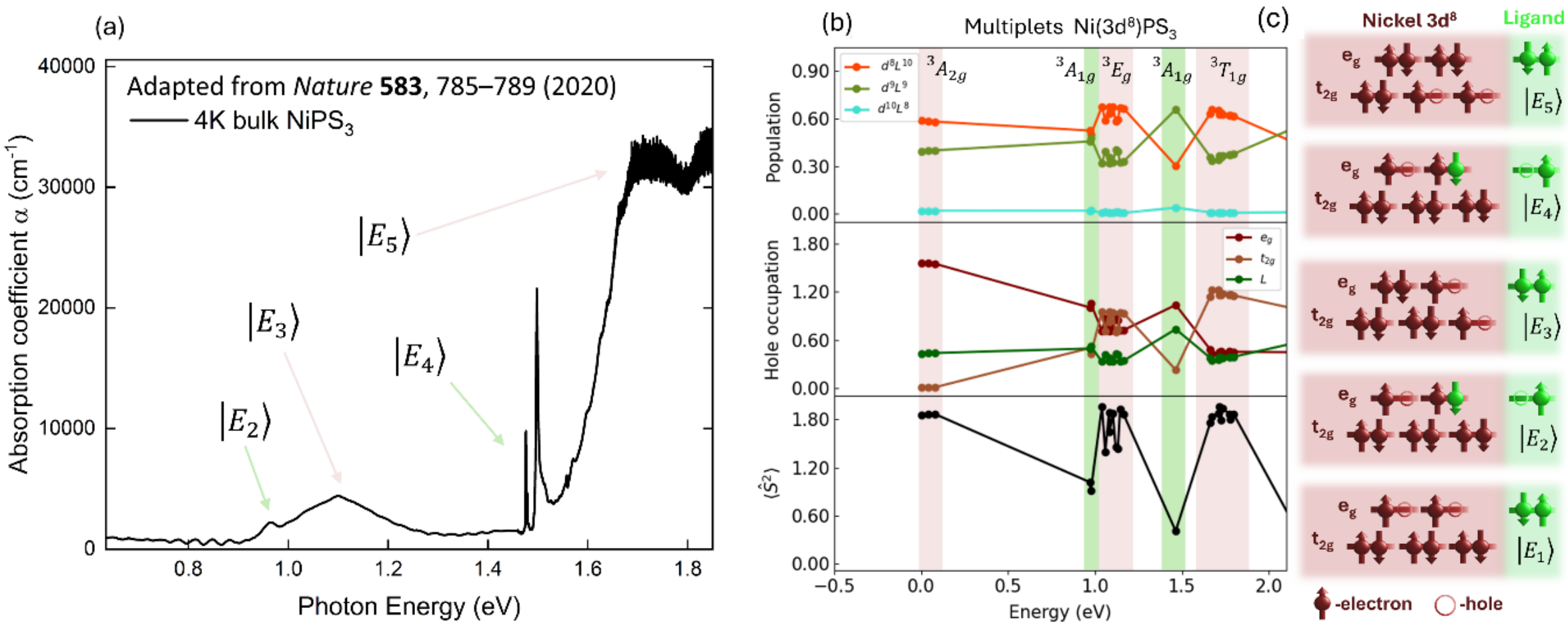

**Figure 8.** (a) Experimental optical absorption spectrum of bulk $Ni_2P_2S_6$ at 4 K, adapted from Ref. [10], with the calculated multiplet energies from panel (b) superimposed. (b) Energies, hole occupancies, and $\langle \hat{S}^2 \rangle$ values of the lowest Ni ($3d^8$) local multiplets obtained for the $NiS_6$ cluster using the single-impurity Anderson model, cf. [56]. (c) Schematic orbital configurations of the five lowest excited multiplets $|E_1\rangle - |E_5\rangle$, highlighting the redistribution of $e_g$, $t_{2g}$ and ligand-hole character across the excitation manifold.

## Tables:

**Table I.** Summary of Raman features observed in bulk and exfoliated $Ni_2P_2S_6$ at $T = 4K$ under resonant $\lambda_{exc} = 633$ nm excitation. The labels P1-P22 denote Raman peaks assigned to intrinsic $Ni_2P_2S_6$ spectral features, while S1-S3 denote additional features attributed to flake-only, symmetry-activated, or substrate-related contributions. $\nu_{bulk}$ is the peak position measured in bulk $Ni_2P_2S_6$, and $\hbar\omega$ is the corresponding Raman energy. The column $\nu_{20/15/10nm}$ lists the peak positions measured for exfoliated flakes with thicknesses of 20, 15, and 10 nm, respectively. The assignment column indicates the proposed dominant scattering process: 1R denotes first-order Raman scattering by a zone-center phonon, 2R denotes second-order or two-phonon Raman scattering, and DR denotes a double-resonance-activated process involving finite-wave-vector phonons. In the last column, $A_g$ and $B_g$ denote Raman-active irreducible representations of the monoclinic crystal symmetry, whereas expressions such as 2×P6, P6+P9, or P6(q)+P6(−q) indicate overtone or combination processes involving the corresponding first-order modes. The symbol "–" denotes that no irreducible representation or phonon-combination label is assigned.

| Label | $\nu_{bulk}$ (cm$^{-1}$) | $\hbar\omega$ (meV) | $\nu_{20/15/10}$ nm | Assignment/dominant process | Irrep/combo |
|---|---|---|---|---|---|
| P1 | 133.1 | 16.50 | 134.1/131.2/134.9 | 1R: low-frequency lattice vibration | $A_g$ |
| P2 | 178.7 | 22.15 | 180.9/180.4/179.9 | 1R: bending-type Γ phonon | $B_g$ |
| S1 | 204.7 | 25.38 | 204.7/205.5/205.9 | Flake-only feature (symmetry-activated) | – |
| P3 | 237.0 | 29.39 | 239.3/239.4/238.7 | 1R: mixed Ni–S vibration | $A_g$ |
| P4 | 255.3 | 31.66 | 256.8/256.8/256.0 | 1R: strong symmetric stretching mode | $A_g$ |
| P5 | 281.9 | 34.95 | 274.2/274.1/274.2 | 1R: interlayer-sensitive vibration | $B_g$ |
| S2 | 304.8 | 37.79 | 304.8/304.7/304.7 | Flake-only feature (symmetry-activated) | – |
| P6 | 385.2 | 47.76 | 386.4/386.6/386.6 | 1R: dominant fingerprint phonon | $A_g$ |
| P7 | 441.8 | 54.78 | 444.5/443.8/440.5 | DR-activated single phonon (finite-q) | DR |
| P9 | 590.3 | 73.19 | 591.5/591.5/591.5 | 1R: high-frequency internal vibration | $A_g$ |
| P10 | 594.3 | 73.69 | 596.2/596.1/593.0 | 1R: split component of P9 | $A_g$ |
| S3 | 622.3 | 77.15 | 622.3/620.8/620.9 | Flake-only optical feature | – |
| P11 | 730.7 | 90.60 | 730.7/730.5/730.4 | 2R: weak combination band | ~ P6 + acoustic |
| P12 | 755.3 | 93.65 | 755.5/756.1/755.6 | 2R: overtone | 2 × P6 |
| P13 | 773.8 | 95.94 | 774.2/774.4/773.6 | 2R: overtone (dispersive) | 2 × P6 |
| P14 | 788.8 | 97.79 | 790.1/787.4/789.5 | 2R: combination, resonantly enhanced | P6(q) +P6(−q) |
| P15 | 808.1 | 100.19 | 807.8/806.5/807.9 | 2R: dominant double-resonant process | P6(q) +P6 (−q) |
| P16 | 818.7 | 101.51 | 819.2/819.6/819.1 | 2R: double-resonant tail | P6(q) +P6(−q) |
| P17 | 841.2 | 104.30 | 835.8/833.0/832.1 | 2R: strong combination band | P6(q) +P6(−q) |
| P18 | 866.9 | 107.48 | 866.9/869.1/865.9 | 2R: upper-energy continuation | P6(q) +P6(−q) |
| P19 | 1131.4 | 140.28 | 1137.0/1130.1/1130.0 | 2R: combination of dominant modes | P6 + P9 |
| P20 | 1169.9 | 145.05 | 1167.0/1166.0/1164.2 | 2R: overtone (lower component) | 2 × P9 |
| P21 | 1190.4 | 147.60 | 1191.0/1190.9/1190.4 | 2R: overtone (upper component) | 2 × P9 |
| P22 | 1589.5 | 197.08 | 1594.2/1595.4/1596.1 | High-energy feature: higher-order scattering | – |

# Supplementary Information

## Resonantly Enhanced Multiphonon Scattering and Local Orbital-Phonon Coupling in Bulk and Thin Flakes of 2D Single Crystals of Antiferromagnetic $(Ni_xFe_{1-x})_2P_2S_6$

Nasaru Khan[1*], Miłosz Rybak[2], Yuliia Shemerliuk[3], Sebastian Selter[3], Bernd Büchner[3,4], Saicharan Aswartham[3,5], Krzysztof Wohlfeld[6] and Pradeep Kumar[1,†]

[1] *School of Physical Sciences, Indian Institute of Technology Mandi, Mandi-175005, India*
[2]*Department of Semiconductor Materials Engineering, Faculty of Fundamental Problems of Technology, Wrocław University of Science and Technology Wybrzeże Wyspiańskiego 27, 50-370 Wrocław, Poland*
[3]*Leibniz-Institute for Solid-state and Materials Research, IFW-Dresden, 01069 Dresden, Germany*
[4]*Institute of Solid State and Materials Physics and Würzburg-Dresden Cluster of Excellence ct.qmat, Technische Universität Dresden, 01062 Dresden, Germany*
[5]*International Research Centre MagTop, Institute of Physics, Polish Academy of Sciences, al. Lotników 32/46, 02-668 Warsaw, Poland*
[6]*Institute of Theoretical Physics, Faculty of Physics, University of Warsaw, Pasteura St. 5, 02-093 Warsaw, Poland*

[*]nasarukhan736@gmail.com

[†]pkumar@iitmandi.ac.in

### Section A. Raman scattering in $(Ni_xFe_{1-x})_2P_2S_6$

The unit cells of $Ni_2P_2S_6$ and $Fe_2P_2S_6$ yield a total of 30 phonon modes at the Γ-point of the Brillouin zone, classified according to their irreducible representations as $8A_g + 6A_u + 7B_g + 9B_u$. Of these, 15 modes are Raman active ($\Gamma_{Raman} = 8A_g + 7B_g$), 12 ($\Gamma_{IR} = 5A_u + 7B_u$), are infrared active, and the remaining 3 ($\Gamma_{Acoustic} = A_u + 2B_u$) correspond to acoustic phonons. Supplementary Figure S5 presents the Raman spectra of the $(Ni_xFe_{1-x})_2P_2S_6$ series for compositions *x* = 1.0, 0.7, 0.5, 0.3, and 0.0, measured in the spectral range of 50-1700 $cm^{-1}$ using a 532 nm excitation laser. Insets in Figure S5 display the fitted Raman spectra for the same single-crystal series in the higher-frequency region of 700-1700 $cm^{-1}$. For clarity, the observed modes are labeled as P1-P22, G1-G4, R1-R2, and Q1. In the case of bulk $Ni_2P_2S_6$, 22 distinct Raman modes are observed, of which 10 correspond to first-order phonon modes. Additionally, the P8 mode exhibits a Fano lineshape when excited with the 532 nm laser. For the doped compositions, the number of observed Raman features increases: 28 for $(Ni_{0.7}Fe_{0.3})_2P_2S_6$, 29 for $(Ni_{0.5}Fe_{0.5})_2P_2S_6$, and 28 for $(Ni_{0.3}Fe_{0.7})_2P_2S_6$. Among these, 14 modes are identified as first-order phonon modes across all doped samples. The origins of these first-order modes, along with their temperature-dependent self-energy behavior, have been previously reported in Ref. [8].

We now turn our attention to the higher-order Raman modes. The high-frequency region is characterized by two prominent clusters of peaks located near ~ 800 $cm^{-1}$ and ~1200 $cm^{-1}$. In $Ni_2P_2S_6$, the spectral range between 700-900 $cm^{-1}$ is particularly rich, containing multiple features that can be well-fitted using 8 Lorentzian line shapes. A subset of these modes may be attributed to second-order overtones of phonons near ~ 385 $cm^{-1}$. This spectral structure is also evident in the doped systems; however, it is entirely absent in $Fe_2P_2S_6$. Notably, as the Ni content decreases, the intensity and number of features within this cluster systematically diminish. For example, in $(Ni_{0.7}Fe_{0.3})_2P_2S_6$, two modes (P11 and P14) from the original cluster vanish. In the more heavily doped samples $(Ni_{0.5}Fe_{0.5})_2P_2S_6$ and $(Ni_{0.3}Fe_{0.7})_2P_2S_6$, an additional mode (P12) disappears. A new weak mode near ~ 704 $cm^{-1}$ (labeled as Q1) emerges in these compositions. Overall, the progressive weakening and eventual disappearance of the 700-900 $cm^{-1}$ cluster with reduced Ni content strongly indicates that this spectral feature originates from Ni-associated lattice dynamics or electronic interactions, underscoring the element-specific nature of these higher-order modes.

Another notable spectral feature emerges in the high-frequency range of 1000-1200 $cm^{-1}$, where a distinct cluster of peaks is observed. This cluster can be deconvoluted into 3 Lorentzian components, labelled as modes P19, P20, and P21. Among these, P21 is the most prominent and consistently appears in all samples containing Ni. P20, acting as a shoulder to P21, is likewise present across all Ni-containing compositions. In contrast, mode P19 progressively weakens with decreasing Ni content and is absent in $(Ni_{0.5}Fe_{0.5})_2P_2S_6$ and $(Ni_{0.3}Fe_{0.7})_2P_2S_6$. Strikingly, the entire ~ 1200 $cm^{-1}$ cluster disappears for $Fe_2P_2S_6$, reaffirming that this spectral feature originates from Ni-specific interactions likely tied to orbiton-phonon hybrid modes.

Supplementary Figure S3 presents the temperature evolution of some of the prominent higher-order Raman modes for $(Ni_{0.7}Fe_{0.3})_2P_2S_6$, $(Ni_{0.5}Fe_{0.5})_2P_2S_6$, and $(Ni_{0.3}Fe_{0.7})_2P_2S_6$. The Peak frequency and FWHM of these modes show the slope anomaly in the vicinity of temperature where these systems enter into antiferromagnetic (AFM) phase from paramagnetic phase. These anomalies establish a clear correlation between the self-energy parameters of the Raman modes and the onset of AFM ordering. In $(Ni_{0.5}Fe_{0.5})_2P_2S_6$ and $(Ni_{0.3}Fe_{0.7})_2P_2S_6$, the frequencies of modes P12, P16, P20, and P21 show a conventional decrease with increasing temperature, as expected for phonon modes. Similarly, in $(Ni_{0.7}Fe_{0.3})_2P_2S_6$, modes P15, P16, P20, and P21 follow this normal softening trend with increasing temperature. Notably, the AFM transition temperatures for $(Ni_{0.7}Fe_{0.3})_2P_2S_6$, $(Ni_{0.5}Fe_{0.5})_2P_2S_6$, and $(Ni_{0.3}Fe_{0.7})_2P_2S_6$ are ~ 135 K, ~ 130 K, and ~ 120 K, respectively [8, 65]. In all three cases, we observe pronounced deviations in the temperature dependence of both frequency and FWHM near the AFM transition points. These slope changes signal the emergence of long-range magnetic order and underscore the sensitivity of higher-order Raman modes to magnetic phase transitions. Thus, the evolution of self-energy parameters in these doped systems provides evidence of spin-lattice coupling across the AFM transition.

**Section B. Additional first-order phonon modes in few layered $Ni_2P_2S_6$**

In the few-layered $Ni_2P_2S_6$ systems (20 nm, 15 nm, and 10 nm flakes), three additional Raman modes labelled as S1, S2, and S3 are observed. The S1 mode appears at 204.7 $cm^{-1}$, 205.0 $cm^{-1}$, and 205.9 $cm^{-1}$ for 20 nm, 15 nm, and 10 nm flakes, respectively, consistent with earlier reports for few-layered $Ni_2P_2S_6$ [66]. The S2 mode is found at 304.8 $cm^{-1}$ (20 nm), 304.7 $cm^{-1}$ (15 nm), and 304.7 $cm^{-1}$ (10 nm), while the S3 mode occurs at 622.3 $cm^{-1}$, 620.8 $cm^{-1}$, and 620.9 $cm^{-1}$ for the corresponding flake thicknesses. To further analyse these modes, we extracted their self-energy parameters i.e. mode frequency, FWHM, and the intensity for the 10 nm flake using Lorentzian fitting. Supplementary Figure S4 illustrates the temperature evolution of these parameters for modes S1-S2 Following observations can be made (i) Frequency of these modes exhibit typical phonon softening behaviour, i.e., frequency decreases with increasing temperature. However, a distinct slope change is observed near the antiferromagnetic transition temperature ($T_N$) ~ 150 K. (ii) FWHM S3 mode shows a normal increase in linewidth with temperature, along with a noticeable slope change near $T_N$. (iii) $S_2$ mode displays a non-monotonic trend - FWHM increases from 4 K to $T_N$ and then decreases up to 330 K. (iv) Interestingly, $S_1$ mode exhibits anomalous behavior, with FWHM decreasing as temperature increases, and again showing a transition near $T_N$. (v) Intensity of modes S1 and S3 exhibit normal behaviour as it decreases with increasing temperature. In contrast, S2 displays an unconventional trend - its intensity increases with temperature. Notably, all three modes show distinct changes in intensity slope around $T_N$, underscoring their sensitivity to magnetic ordering.

The appearance of the additional Raman modes S1-S3 in few-layered $Ni_2P_2S_6$, which are absent in bulk counterparts raises intriguing possibilities regarding their physical origin. Based on our observations, here are the most plausible possibilities. Reduction in dimensionality from bulk to thin flakes may breaks inversion symmetry, at least locally if not globally and may modifies crystal field environments, potentially activating otherwise forbidden or silent modes. These modes may arise from Davydov splitting, where phonon modes split in few-layer limits due to modified interlayer interactions. However, more experimental studies are required to shed light on the origin of these modes.

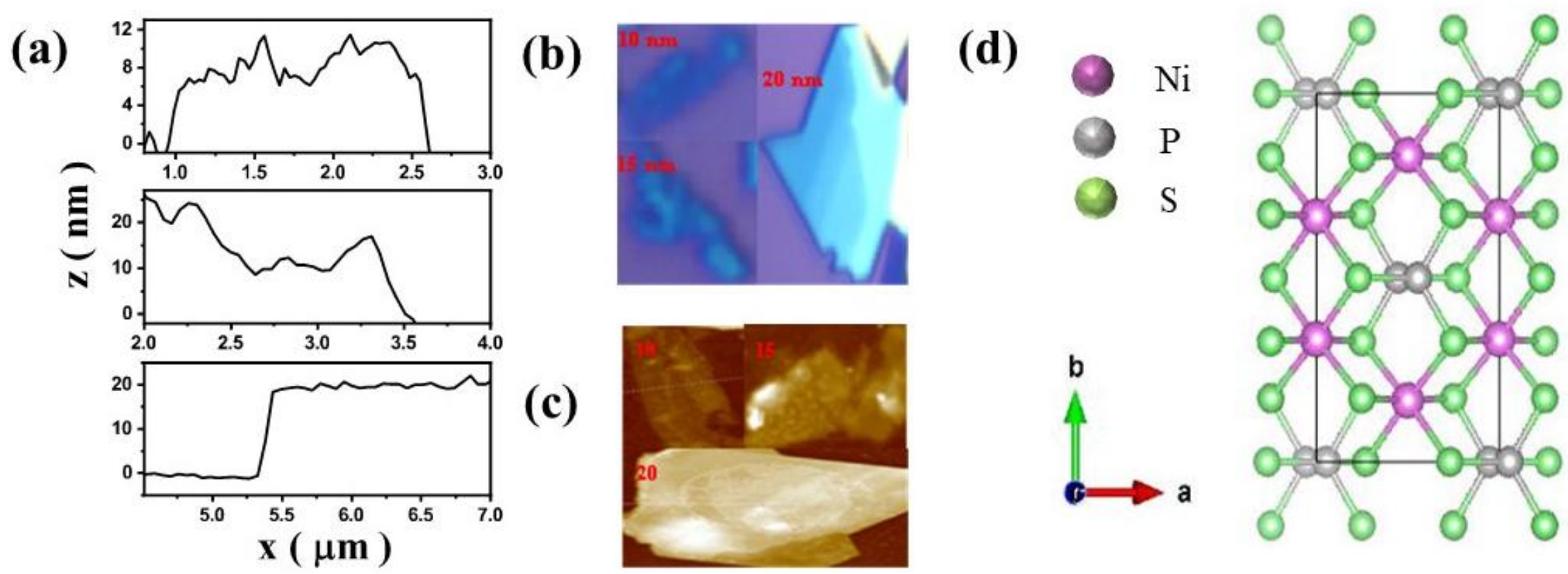

**Figure S1. (a)** Height profile of exfoliated $Ni_2P_2S_6$ nanosheets, with **z** representing the flake thickness determined using AFM. **(b)** Optical microscopy images of $Ni_2P_2S_6$ flakes deposited on a $SiO_2$/Si substrate. **(c)** Tapping-mode AFM topography images showcasing the surface morphology of the exfoliated flakes. **(d)** Crystallographic structure of $Ni_2P_2S_6$ projected along the *ab* plane, highlighting atomic arrangement within the layer.

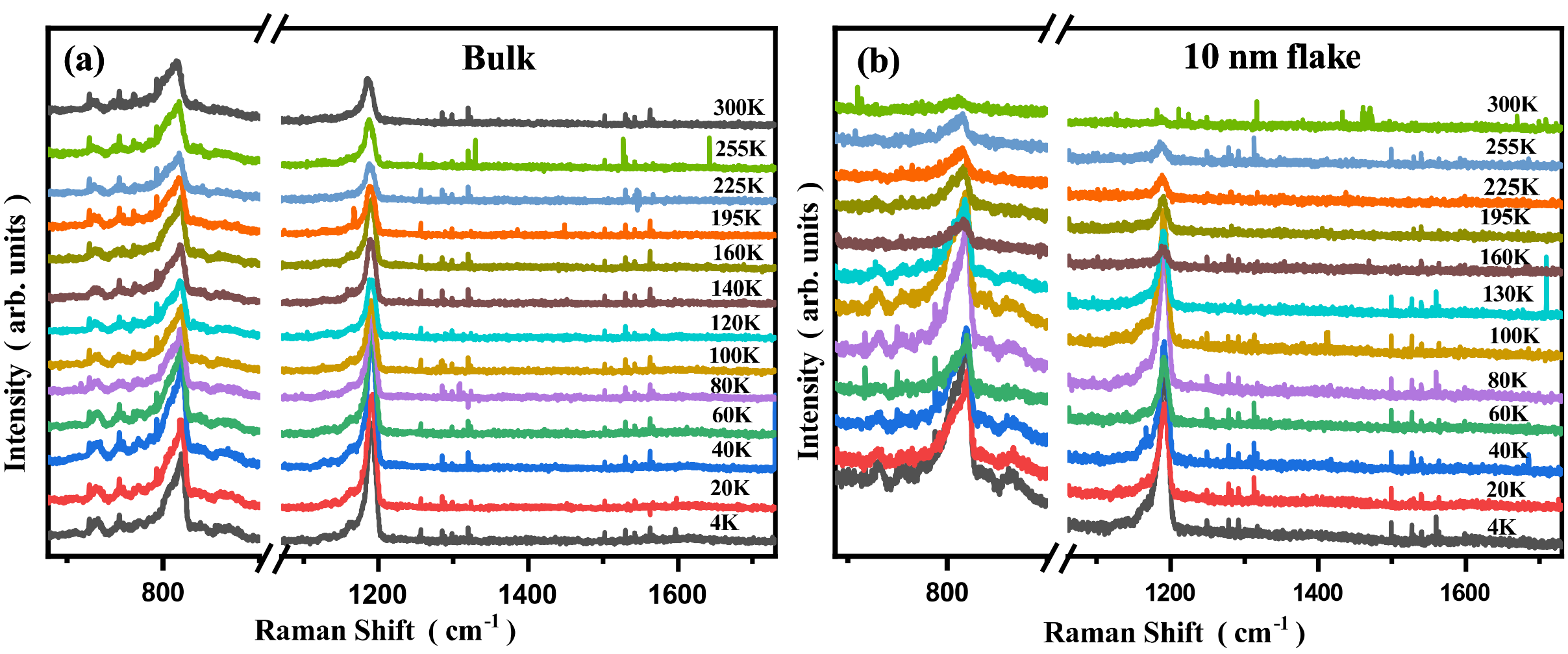

**Figure S2.** Raman spectra of **(a)** bulk and **(b)** 10 nm $Ni_2P_2S_6$ measured at various temperatures using a 633 nm excitation laser.

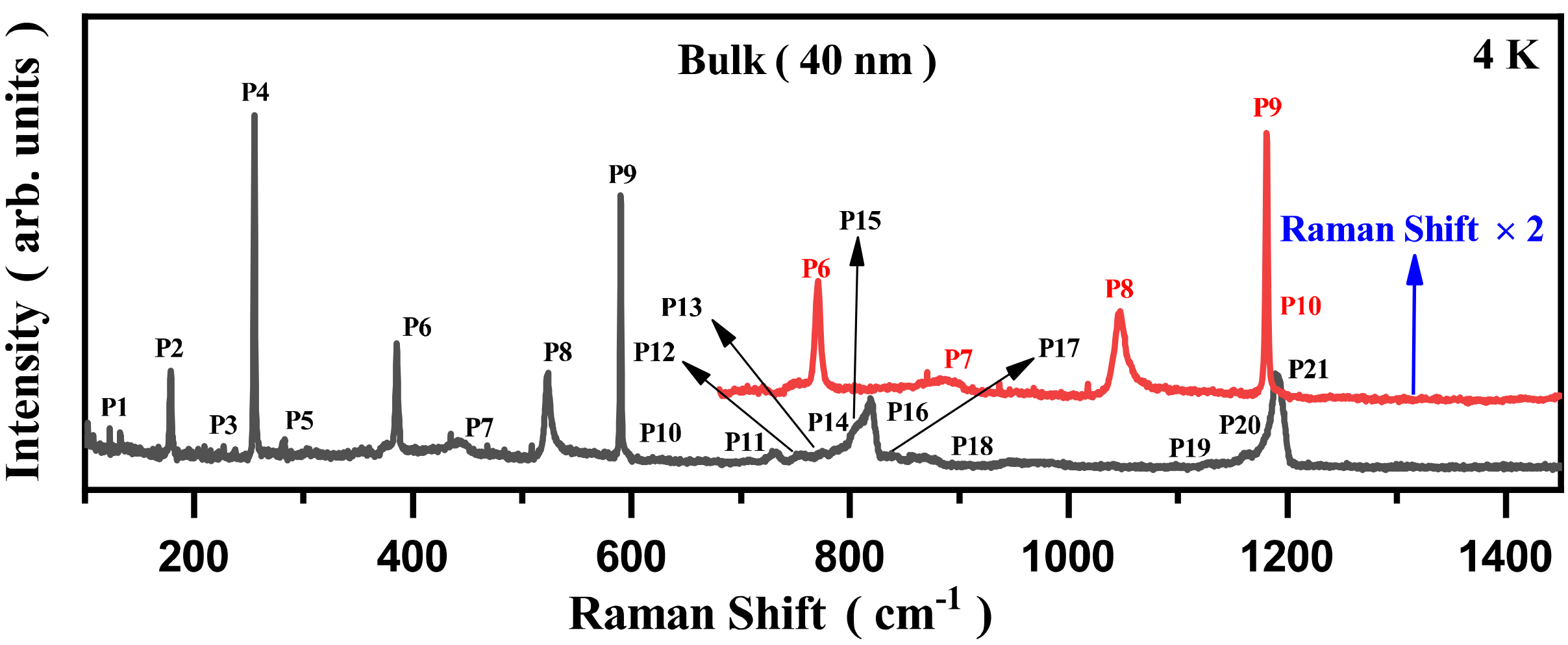

**Figure S3.** Raw Raman spectrum of bulk $Ni_2P_2S_6$. To aid in identifying potential two-phonon excitations, the spectrum is also plotted as a function of twice the Raman shift (indicated in red).

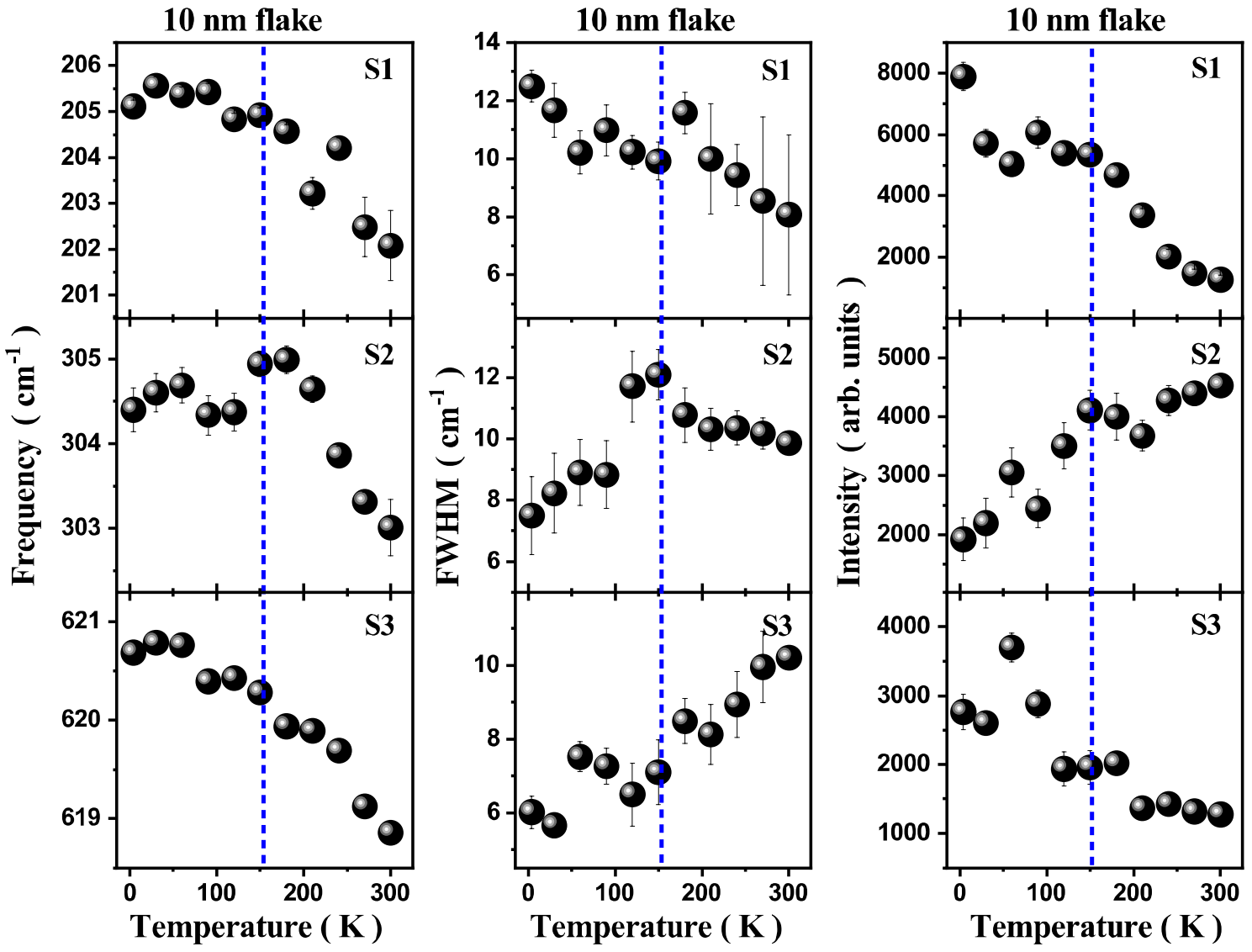

**Figure S4.** Temperature-dependent evolution of the peak frequency, FWHM, and intensity for modes S1-S3 in the 10 nm $Ni_2P_2S_6$ flake.

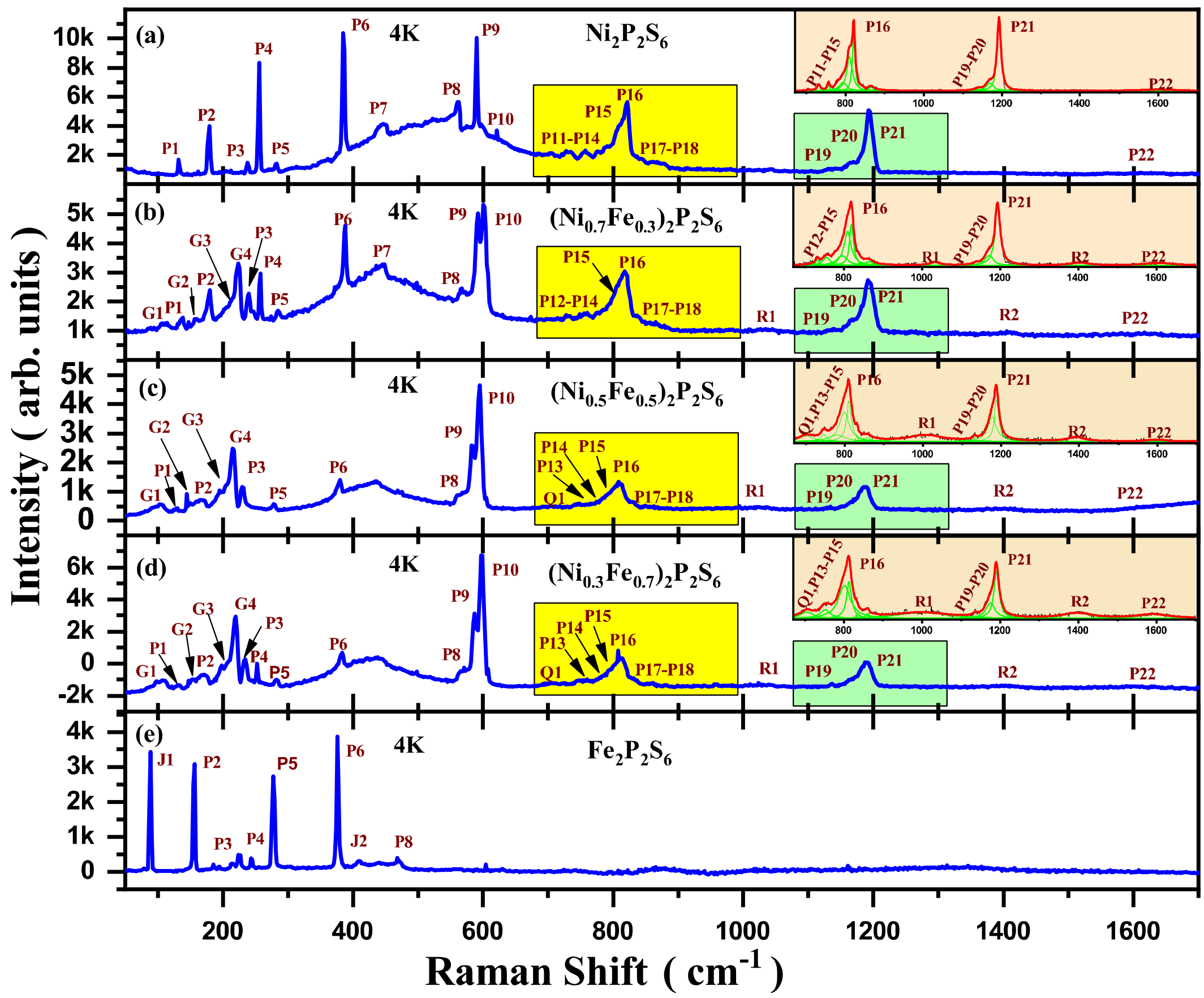

**Figure S5.** Raw Raman spectra recorded at 4 K using a 532 nm excitation laser in the spectral range of 50-1700 cm$^{-1}$ for (a) $Ni_2P_2S_6$, (b) $(Ni_{0.7}Fe_{0.3})_2P_2S_6$, (c) $(Ni_{0.5}Fe_{0.3})_2P_2S_6$, (d) $Ni_{0.7}Fe_{0.3})_2P_2S_6$, and (e) $Fe_2P_2S_6$. The yellow shaded region marks a cluster of peaks fitted with multiple Lorentzian line shapes, while the green shaded region highlights a distinct cluster of three Lorentzian modes. Raman-active features are labelled as P11-P22. Insets display the fitted spectra for the Ni-containing samples within the 670-1700 cm$^{-1}$ range, emphasizing the detailed peak decomposition in the higher-frequency region.

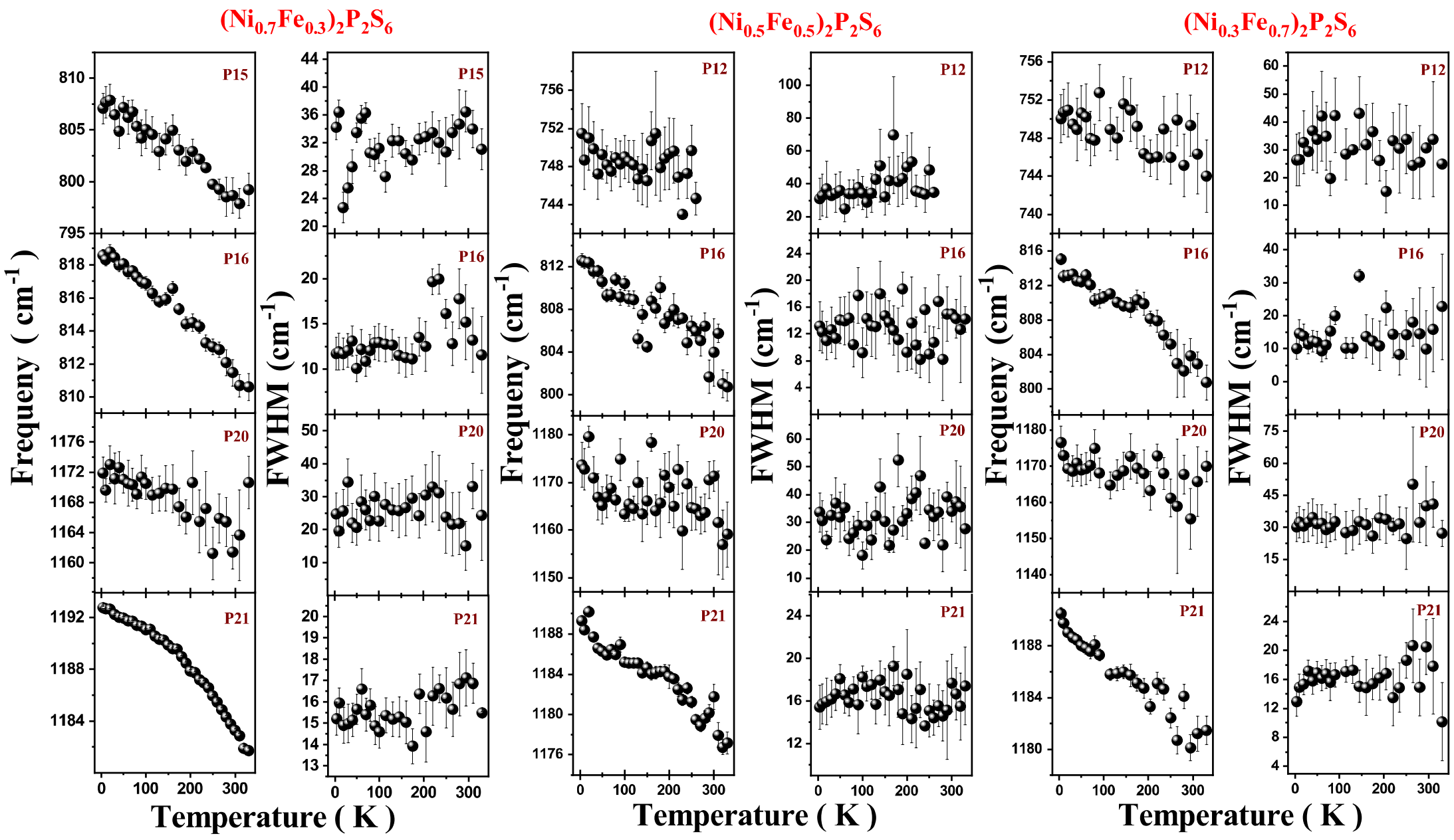

**Figure S6.** Temperature-dependent evolution of frequency and FWHM for selected higher-order Raman modes in **(a)** $(Ni_{0.7}Fe_{0.3})_2P_2S_6$, **(b)** $(Ni_{0.5}Fe_{0.5})_2P_2S_6$, and **(c)** $(Ni_{0.3}Fe_{0.7})_2P_2S_6$.

**TABLE S1**: List of experimentally observed Raman modes at 4 K for bulk and exfoliated $Ni_2P_2S_6$ flakes with thicknesses of 20 nm, 15 nm, and 10 nm.

| **Mode** | **Mode frequency at 4 K ($\omega_0$)** | | | |
|---|---|---|---|---|
| | **Bulk** | **20 nm Flake** | **15 nm Flake** | **10 nm Flake** |
| P1 | 133.1 ± 1.8 | 134.1 ± 0.2 | 131.2 ± 0.1 | 134.9 ± 0.3 |
| P2 | 178.7 ± 0.1 | 180.9 ± 0.1 | 180.4 ± 0.1 | 179.9 ± 0.1 |
| S1 | | 204.7 ± 1.0 | 205.5 ± 0.5 | 205.9 ± 0.2 |
| P3 | 237.0 ± 3.6 | 239.3 ± 0.4 | 239.4 ± 0.4 | 238.7 ± 0.4 |
| P4 | 255.3 ± 0.1 | 256.8 ± 0.1 | 256.8 ± 0.1 | 256.0 ± 0.1 |
| P5 | 281.9 ± 0.4 | 274.2 ± 0.9 | 274.1 ± 1.1 | 274.2 ± 0.3 |
| S2 | | 304.8 ± 0.2 | 304.7 ± 0.9 | 304.7 ± 0.6 |
| P6 | 385.2 ± 0.1 | 386.4 ± 0.1 | 386.6 ± 0.1 | 386.6 ± 0.1 |
| P7 | 441.8 ± 1.2 | 444.5 ± 0.4 | 443.8 ± 1.5 | 440.5 ± 0.6 |
| P8 | | | | |
| P9 | 590.3 ±0.1 | 591.5 ± 0.4 | 591.5 ± 0.1 | 591.5 ± 0.1 |
| P10 | 594.3 ± 0.1 | 596.2 ± 0.1 | 596.1 ± 0.3 | 593.0 ± 0.7 |
| S3 | | 622.3 ± 1.8 | 620.8 ± 0.6 | 620.9 ± 1.6 |
| P11 | 730.7 ± 0.2 | 730.7 ± 0.2 | 730.5 ± 0.3 | 730.4 ± 0.4 |
| P12 | 755.3 ± 0.3 | 755.5 ± 0.3 | 756.1 ± 0.4 | 755.6 ± 0.5 |
| P13 | 773.8 ± 0.4 | 774.2 ± 0.4 | 774.4 ± 0.3 | 773.6 ± 0.6 |
| P14 | 788.8 ± 0.7 | 790.1 ± 0.9 | 787.4 ± 0.3 | 789.5 ± 1.6 |
| P15 | 808.1 ± 0.2 | 807.8 ± 0.2 | 806.5 ± 0.1 | 807.9 ± 0.3 |
| P16 | 818.7 ± 0.1 | 819.2 ± 0.1 | 819.6 ± 0.1 | 819.1 ± 0.1 |
| P17 | 841.2 ± 0.4 | 835.8 ± 0.4 | 833.0 ± 0.3 | 832.1 ± 0.5 |
| P17A | | 843.1 ± 0.3 | 842.5 ± 0.3 | 842.2 ± 0.5 |
| P18 | 866.9 ± 0.5 | 866.9 ± 0.3 | 869.1 ± 0.2 | 865.9 ± 0.4 |
| P19 | 1131.4 ± 1.6 | 1137.0 ± 2.8 | 1130.1 ± 2.3 | 1130.0 ± 2.5 |
| P20 | 1169.9 ± 0.6 | 1167.0 ± 0.2 | 1166.0 ± 0.1 | 1164.2 ± 0.2 |
| P21 | 1190.4 ± 0.1 | 1191.0 ± 0.1 | 1190.9 ± 0.1 | 1190.4 ± 0.1 |
| P22 | 1589.5 ± 0.5 | 1594.2 ± 0.2 | 15945.4 ± 0.4 | 1596.1± 0.2 |

**TABLE S2:** Atoms and the corresponding Wyckoff positions in the unit cell and irreducible representations of the phonon modes for the monoclinic (point group $(C_{2h})$ and space group ( $(C2/m,\ \#12)$) $Ni_2P_2S_6$ at the $\Gamma$ point. Irreducible representation $\Gamma_{Total}$, $\Gamma_{Raman}$, $\Gamma_{IR}$, $\Gamma_{Acoustic}$ corresponds to total, Raman, Infrared and acoustic phonon modes representation at the $\Gamma$ point, respectively. $R_{A_g}$ and $R_{B_g}$ are the Raman tensor for the Raman active $A_g$ and $B_g$ phonon mode, respectively.

| Atom | Type | Site | $\Gamma$ point phonon mode decomposition | Raman Tensors |
|---|---|---|---|---|
| Ni1 | Ni | 4g | $A_g + A_u + 2B_g + 2B_u$ | $R_{A_g} = \begin{pmatrix} a & 0 & d \\ 0 & b & 0 \\ d & 0 & c \end{pmatrix}$ |
| P1 | P | 4i | $2A_g + A_u + B_g + 2B_u$ | |
| S1 | S | 4i | $2A_g + A_u + B_g + 2B_u$ | $R_{B_g} = \begin{pmatrix} 0 & e & 0 \\ e & 0 & f \\ 0 & f & 0 \end{pmatrix}$ |
| S2 | S | 8j | $3A_g + 3A_u + 3B_g + 3B_u$ | |
| $\Gamma_{Total} = 8A_g + 6A_u + 7B_g + 9B_u$, $\Gamma_{Raman} = 8A_g + 7B_g$, $\Gamma_{IR} = 5A_u + 7B_u$, $\Gamma_{Acoustic} = A_u + 2B_u$ | | | | |